\documentclass[journal]{IEEEtran}

  \usepackage[nocompress]{cite}
  \usepackage{tcolorbox}
  \usepackage{tikz}
\usepackage{amsmath}

\usetikzlibrary{positioning}

\usepackage{filecontents}
\usepackage{enumitem}

\usepackage{framed}

\usepackage{amssymb}

\usepackage{amsthm}

\usepackage{multirow}
 
\usepackage{caption}
\usepackage{makecell}
\usepackage{colortbl}
\usepackage{pifont}
\usepackage{multicol}
\usepackage{booktabs}
\usepackage{algorithmic}
\usepackage[ruled,vlined]{algorithm2e} 

\usepackage{xspace}

\usepackage{pifont}

\usepackage{float}

\graphicspath{{./figure/}}

\usepackage[colorlinks=true,bookmarks=false,citecolor=blue,urlcolor=blue]{hyperref}

\usepackage{cleveref}

\usepackage{tcolorbox}

\usepackage{graphicx}

\usepackage{subcaption}

\usepackage{cite}

\ifCLASSINFOpdf
\else
\fi
\begin{document}

\title{SSD: A State-based Stealthy Backdoor Attack For IMU/GNSS Navigation System in
UAV Route Planning}
%
%

\author{Zhaoxuan~Wang,~\IEEEmembership{}
Yang ~Li,~\IEEEmembership{Member,~IEEE,}
Jie ~Zhang,~\IEEEmembership{}
Xingshuo  ~Han,~\IEEEmembership{}
Kangbo  ~Liu,~\IEEEmembership{}
Yang  ~Lyu,~\IEEEmembership{}
Yuan  ~Zhou,~\IEEEmembership{}
Tianwei  ~Zhang,~\IEEEmembership{Member,~IEEE,}
and~Quan ~Pan,~\IEEEmembership{Member,~IEEE.}

\thanks{Zhaoxuan Wang is with School of Cybersecurity, Northwestern Polytechnical University, Xi'an 710129, China (e-mail: zxwang@mail.nwpu.edu.cn).}

\thanks{Yang Li, Yang Lyu, Kangbo Liu and Quan Pan are with School of Automation, Northwestern Polytechnical University, Xi'an 710129, China (e-mail:liyangnpu@nwpu.edu.cn; liukangbo@mail.nwpu.edu.cn;  lyu.yang@nwpu.edu.cn quanpan@nwpu.edu.cn).}

\thanks{Jie Zhang is with CFAR and IHPC, Agency for Science, Technology and Research, Singapore. e-mail: (zhang\_jie@cfar.a-star.edu.sg).}

\thanks{Xingshuo Han and Tianwei Zhang are with College of Computing and Data Science, Nanyang Technological University, Singapore 639798 (e-mail: xingshuo001@e.ntu.edu.sg; tianwei.zhang@ntu.edu.sg).}


\thanks{Yuan Zhou is with School of Computer Science and Technology, Zhejiang Sci-Tech University, Zhejiang 310018, China (email: yuanzhou@zstu.edu.cn).}

\thanks{Corresponding author: Yang Li.}

}

\markboth{}
{Shell \MakeLowercase{\textit{et al.}}: Bare Demo of IEEEtran.cls for IEEE Journals}
%


\maketitle

\begin{abstract}
Unmanned aerial vehicles (UAVs) are increasingly employed to perform high-risk tasks that require minimal human intervention. However, they face escalating cybersecurity threats, particularly from GNSS spoofing attacks. While previous studies have extensively investigated the impacts of GNSS spoofing on UAVs, few have focused on its effects on specific tasks. Moreover, the influence of UAV motion states on the assessment of cybersecurity risks is often overlooked. To address these gaps, we first provide a detailed evaluation of how motion states affect the effectiveness of network attacks. We demonstrate that nonlinear motion states not only enhance the effectiveness of position spoofing in GNSS spoofing attacks but also reduce the probability of detecting speed-related attacks. Building upon this, we propose a state-triggered backdoor attack method (SSD) to deceive GNSS systems and assess its risk to trajectory planning tasks. Extensive validation of SSD's effectiveness and stealthiness is conducted. Experimental results show that, with appropriately tuned hyperparameters, SSD significantly increases positioning errors and the risk of task failure, while maintaining high stealthy rates across three state-of-the-art detectors.
\end{abstract}


%
\IEEEpeerreviewmaketitle

\section{Introduction}
\begin{figure}[htp]
	\centering
 \centerline{\includegraphics[width=230pt]{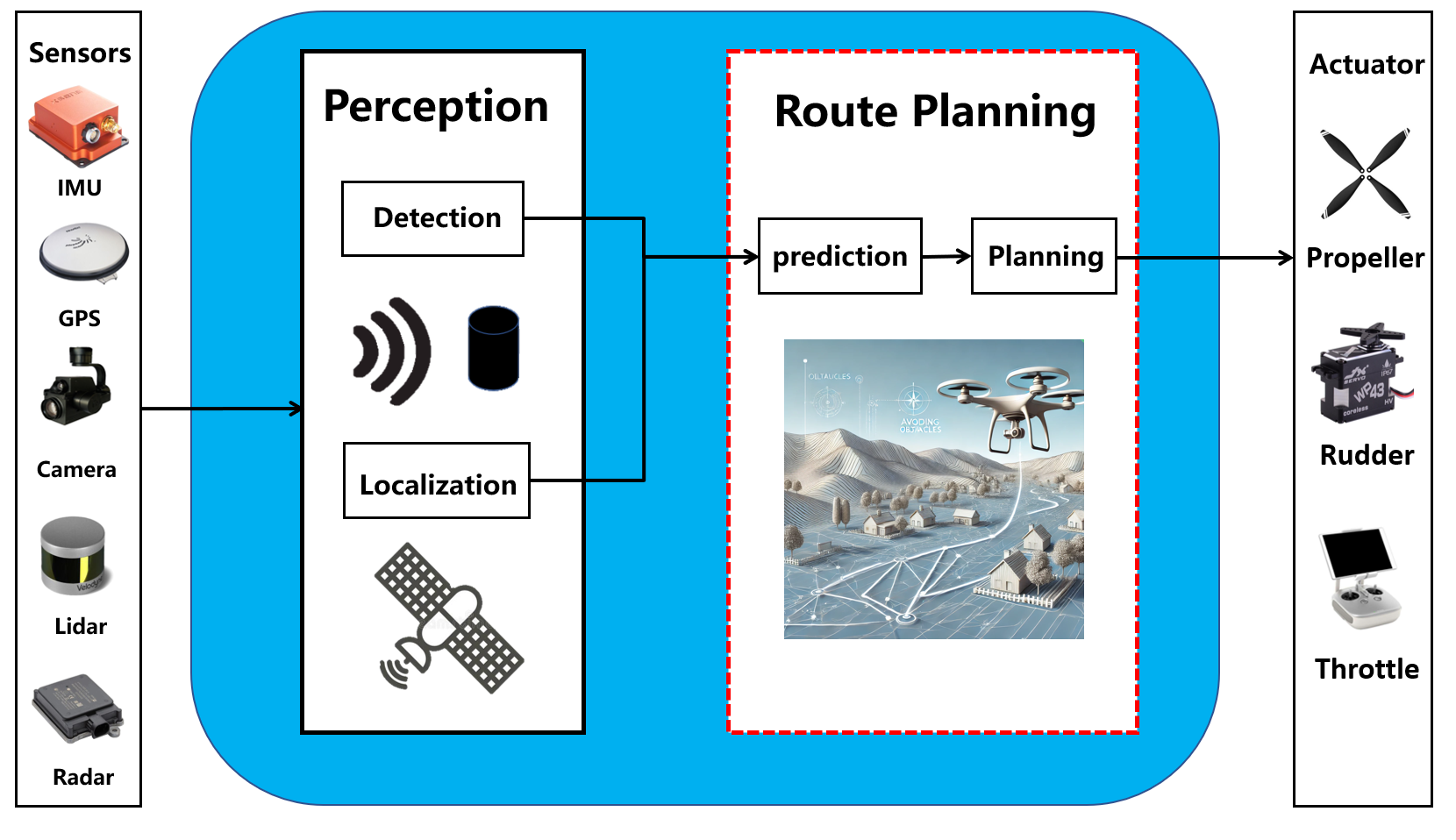}} 
		\caption{The Role of Localization and Route Planing in UAV Autonomous Flight}
		\label{fig:route}
\end{figure}

Unmanned Aerial Vehicles (UAVs) are revolutionizing our understanding of low-altitude flight patterns.
Currently, UAVs are widely used in various military and civilian fields, such as courier delivery, agricultural plant protection, power patrol, firefighting, rescue, and battlefield reconnaissance. 
These increasingly complex application scenarios have necessitated stringent requirements for the autonomous flight capabilities of UAVs. 
As UAVs operate autonomously to execute their missions, the system must precisely determine its global position at a centimeter level.
As illustrated in Fig.~\ref{fig:route}, 
the localization capability is critical to route planning. It ensures the safety of flight and the ability to fulfill its mission, as positioning errors can directly cause the flight to deviate from course or fail to perform its mission.

The Integrated Navigation System (INS) serves as the cornerstone of UAVs, enabling precise positioning. It accomplishes accurate position estimates by integrating data from various sensors. Integrating Inertial Measurement Units (IMUs) and Global Navigation Satellite Systems (GNSS) forms the fundamental and core navigation system in INS. Building on this foundation, researchers usually incorporated additional sensors, such as cameras, lidar, and radar to further improve positioning accuracy in different scenarios or platforms. However, direct reliance on sensor data and communication channels' noise makes INS vulnerable to cyber-attacks \cite{wang2023survey,wei2024survey}. Research has revealed that adversaries can attack INS by using adversarial examples \cite{liu2023rpau} or wireless signal injection \cite{kim2018low} to spoof sensors. Notably, GNSS is a particularly prevalent threat since it forms the foundation of INS. The attacker can leverage low-cost devices to manipulate the position and velocity measurement captured by GNSS.

Previous attacks can be classified into the following categories:(1) \textbf{Direct Attacks}\cite{tang2023gan}
: The adversary directly injects a false signal into the GNSS sensor. (2) \textbf{Stealthy Attacks} \cite{khazraei2024black,ma2024novel}: The adversary takes the detector and corresponding threshold as the constraint and computes an optimization-based payload to bypass the detectors. 

However, these attacks have the following limitations.\textbf{(1) Easily detectable} 
Sensor data fluctuations during UAV attacks are typically significant~\cite{wu2023highly}, prompting residual-based detection methods. However, our experiments show these detectors can almost always detect direct attacks. Moreover, during UAV maneuvers, such as waypoint course changes or formation shifts, the increased fluctuations make attacks even more easily identifiable.
\textbf{(2) Computation efficiency} While stealthy attacks can bypass detectors through constrained optimization models, they typically require extensive matrix operations to solve high-dimensional optimization problems and derive the optimal attack payload in time. These methods introduce significant computational delays in environments with limited resources, especially affecting the attack's real-time performance.
\textbf{(3) Inadequate analysis of dynamic vulnerabilities} 
Research~\cite{shen2020drift} has shown that navigation algorithms in autonomous vehicles are vulnerable to uncertainty during specific periods. UAVs, operating with six degrees of freedom, experience frequent motion changes that affect system stability, especially during GNSS spoofing. Despite this, few studies evaluate how these dynamic motion states influence attack effectiveness.
\textbf{(4) Incomplete assessment methodology} 
Prior studies \cite{kerns2014unmanned,vervisch2017influence} often focus on immediate attack outcomes, such as crashes or path deviations, but fail to assess how attacks affect UAV mission performance and overall efficacy. This leaves a gap in understanding the broader impacts of such attacks on UAV operations.

To overcome these limitations, and gain a deep understanding of the INS vulnerability posed by GNSS attacks, we first provided a detailed interpretable analysis of the relationship between motion states and GNSS spoofing attacks. Specifically, we assess the effects of linear and nonlinear motion states on attack effectiveness. Our findings reveal that changes in motion states amplify the effectiveness of positional attacks and increase the stealthiness of velocity attacks.
We then proposed SSD, a novel stated-based stealthy backdoor attack for GNSS.
Backdoor attacks \cite{chen2022clean,chenbadpre} are a common threat in deep neural networks, where an adversary implants a latent backdoor that remains inactive under normal conditions but is triggered by specific inputs or scenarios, leading to incorrect model predictions. 
Inspired by this concept, we design backdoors for GNSS by using motion state changes as a trigger to initiate staged velocity and positional attacks. In contrast to the stealthy attacks \cite{khazraei2024black,ma2024novel}, this attack mode doesn't need prior knowledge for the detectors and eliminates the need for complex computations. Instead, it is a direct attack that leverages carefully configured parameters and straightforward function calculations to achieve an optimal balance between effectiveness and stealthiness. This makes SSD highly valuable for engineering applications.
Lastly, we selected three representative mission trajectories to assess the effectiveness of SSD. Its performance was compared against existing attack methods. The experimental results demonstrate that SSD maintains detection variables consistently within the threshold range in three classical detectors. 
Furthermore, we introduced evaluation metrics to measure the attack's impact on mission success rates and effectiveness. The experiments also reveal that SSD significantly increases the localization error, thereby effectively disrupting mission completion.

In summary, our contributions are summarized as follows:

\begin{itemize}
    \item We present an interpretable mathematical security study of how motion states influence attack outcomes and demonstrate that UAVs are more vulnerable during maneuvers than in uniform linear flights. We further experimentally prove it.
    
    \item We design SSD, a novel stated-based backdoor attack that utilizes motion state as a trigger to spoof GNSS data. It can execute velocity and positional attacks in stages to simultaneously and covertly attack both states. 
    
    \item We conduct experiments in classic specific mission trajectories and find that SSD can significantly reduce mission completion rates and maintain constant stable stealthiness under attack detection.
    
\end{itemize}

\section{Related Works and Background}
\subsection{UAV Route Planning and Integrated Navigation System}
\label{ins}
UAV route planning involves designing a feasible route from the start point to the destination while meeting all constraints and performance requirements \cite{abdel2024multiobjective}. Effective route planning is crucial for UAVs to complete their missions, and it depends on accurate position estimation. Since UAVs often operate in dynamic and complex environments, they rely on multi-sensor fusion to enhance their ability to perceive the environment. This approach integrates data from various sensors with different modalities and attributes, increasing redundancy and improving reliability in challenging conditions. The integrated navigation system of GNSS and IMU is a typical representative example. Its fusion strategy uses the GNSS data for quantitative updating, IMU data for state prediction, and an optimal estimation framework to achieve accurate positioning in the global coordinate system \cite{qi2002direct,carvalho1997optimal}. 
In GNSS-denied environments, simultaneous localization and mapping (SLAM) that rely on camera~\cite{davison2007monoslam,forster2016svo} and lidar~\cite{zhang2014loam,nguyen2021miliom} are considered more reliable solutions. However, a single sensor alone cannot fully meet the demands for positioning accuracy and response speed. Therefore, combining these sensors with an inertial measurement unit to create IMU/Camera~\cite{qin2018vins,eckenhoff2021mimc} and IMU/Lidar~\cite{xu2022fast} fusion form enables more precise position estimation and improved performance in dynamic environments. This paper focuses on the security analysis of IMU/GNSS INS since it plays a central role in UAVs. It is of generic meaning to study its security.

\subsection{GNSS Spoofing attack}
\label{attack}
Since IMUs are more difficult to manipulate in real-world scenarios, we only discuss GNSS spoofing attacks for IMU/GNSS INS on UAVs. In such attacks, the adversary transmits false location coordinates to the GNSS receiver, thereby concealing the UAV's true location. As a result, unknowingly accepting these false inputs, the navigation system calculates incorrect position information. Specifically, GNSS spoofing includes direct attacks and stealthy attacks. The direct attacks \cite{tang2023gan} can be classified into these categories. (1) \textbf{
Biased Signal Attack}
: These attacks involve adding a bias to the GNSS sensor signals, typically following a uniform distribution. (2) \textbf{Multiplicative Attacks}: In these attacks, the GNSS signals are multiplied by a constant factor, effectively scaling the original signal values. (3) \textbf{Replacement Attacks}: These attacks involve directly replacing the GNSS signals with false or manipulated data. Direct attacks are easy to implement and may lead to disastrous consequences, such as the UAV crashing into obstacles \cite{kerns2014unmanned,vervisch2017influence}. However, most detectors can detect and respond in time. The stealthy attacks~\cite{khazraei2024black,ma2024novel} are diverse, with attackers often aiming to maximize navigation residuals to determine the optimal attack sequence. The design of these attacks may cause the UAV to fall into the malicious attackers’ control~\cite{he2018friendly}. However, it is largely influenced by the detection mechanism and often needs complex computing. SSD integrates the strengths of both approaches. It achieves the same effectiveness as stealthy attacks while requiring low computational resources.

To mitigate this threat, robust countermeasures, such as software analysis \cite{ceccato2021generalized}, cryptography-based authentication \cite{bonior2017implementation} and machine learning-based detections \cite{liang2019detection,han2022ads}, have been implemented to safeguard against GNSS spoofing and ensure the integrity of UAV INS. One important method is multi-sensor fusion \cite{yang2021secure}. As described in section \ref{ins}, it not only provides more accurate estimates for perception and localization. but also enhances the data's trustworthiness, providing greater redundancy for detection and defense in the event of spoofing attacks. For example, Shen et al. \cite{shen2020drift} demonstrated that the effectiveness of constant offset GNSS spoofing attacks is greatly reduced in GNSS/INS/LiDAR fused navigation systems in autonomous driving. However, such multi-sensor fusion strategies also face a period of vulnerability and can not defend against constructed GNSS spoofing against the uncertainty that exists in the fusion algorithm itself. In this paper, we demonstrate a similar phenomenon in UAVs, where changes in motion states significantly amplify the uncertainty of the INS, making it more vulnerable to GNSS attacks. Therefore, we provide an in-depth analysis of the uncertainty and vulnerability and exploit it to design SSD.

\subsection{Threat Model}

\begin{figure}[htp]
	\centering
 \centerline{\includegraphics[width=230pt]{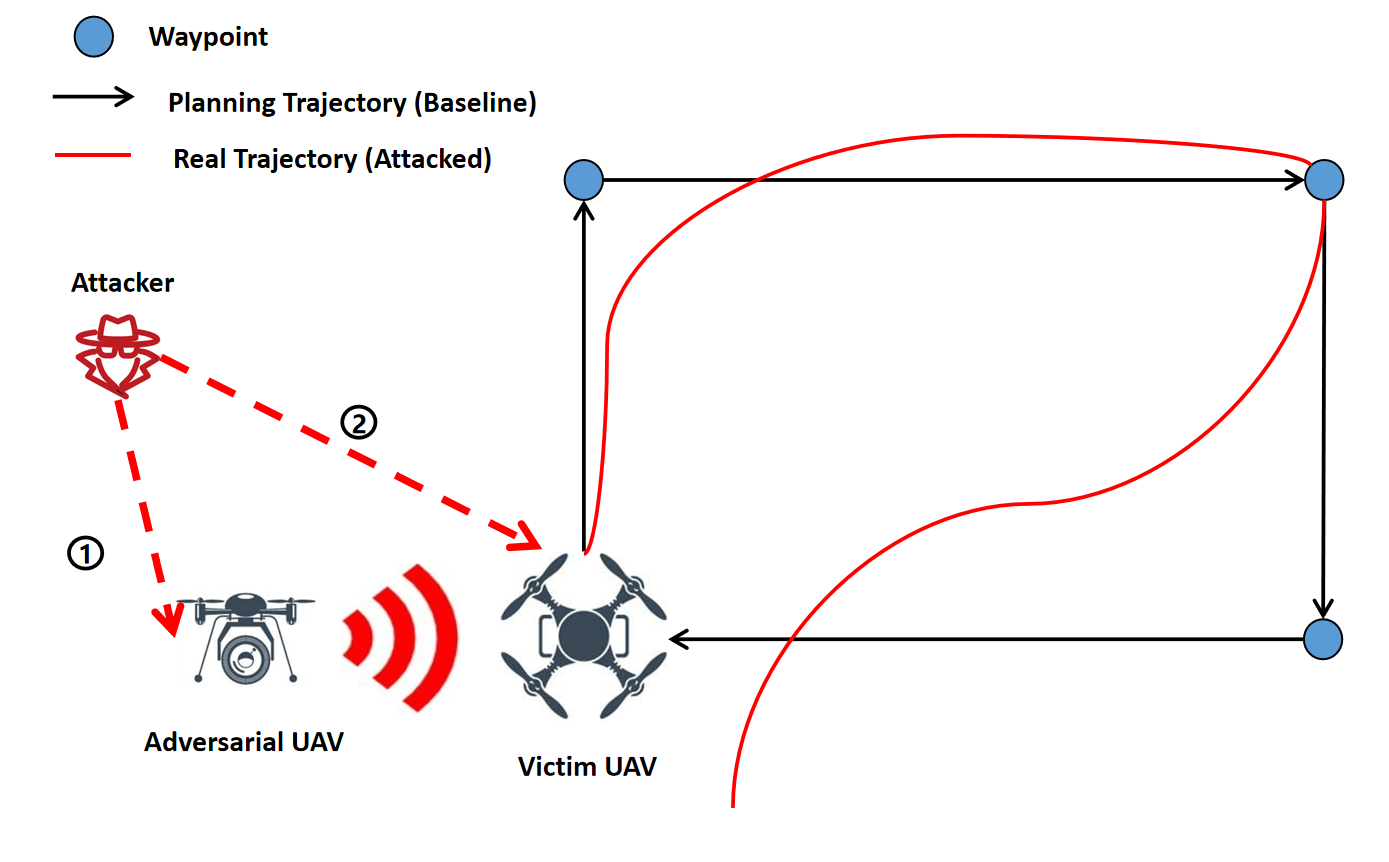}} 
		\caption{Threat Model}
		\label{threat}
\end{figure}

\textbf{Attack Goal} As shown in Figure \ref{threat}, the adversary aims to make the drone deviate significantly from its pre-planned route without triggering the stealth detection threshold. This objective can be formalized as the following optimization problem:

\begin{equation}
    \begin{aligned} \label{P}
&\text{argmax}_{\delta} \quad \sum_{i=0}^{T_a} \mathbf{D^t_i-D^a_i} \\
&\begin{array}{r@{\quad}r@{}l@{\quad}l}
s.t. &\chi_k&\leq \tau \quad \forall k \in \{1,...,T_{a}-1\} \\
\end{array}
\end{aligned}
\end{equation}
where $\mathbf{D^t_i}$ and $\mathbf{D^a_i}$ denote the normal trajectory and the attacked trajectory, respectively. $\delta$ is the attack payload. $\chi_k$ is the in detection statistics and $\tau$ is the threshold of detectors.

\textbf{Attack Scenario} As shown in Fig. \ref{threat}, an attacker can launch an attack in two ways: \ding{172} The attacker can use a UAV to fly alongside the victim's UAV. He can transmit legitimate GNSS signals completely using wireless attack devices such as software-defined radios (SDR). \ding{173} The attacker could also inject specific backdoors \cite{IRAHUL} or viruses~\cite{dji} into the UAV by supply chain attacks, which could be used to monitor its dynamics and induce a false GNSS position \cite{xu2021novel}.

\textbf{Attacker's Capability}
1) The attackers need white-box access to obtain the victim's navigation algorithms and corresponding parameters. They can get this knowledge through open-source channels since most UAVs use standardized open-source navigation algorithms~\cite{ardupilot,px4}. Also, the adversarial can use reverse engineering to access the victim's knowledge. 2) The attacker can obtain the victim's motion state, such as position and velocity. This can be achieved by monitoring UAVs using an additional GPS module or auxiliary object detection and tracking devices. 3) As IMU data is less likely to be accessed and used, an attacker can only modify the position and velocity since GNSS measurements only provide position and velocity to the UAV.

\section{Preliminary}
\subsection{IMU/GNSS Integrated Navigation System}
\label{cns}
 In UAVs, IMU and GNSS are often combined for highly accurate and robust navigation and positioning. IMU provides data from accelerometers and gyroscopes, which measure the acceleration and angular velocity of the UAVs. At the same time, GNSS determines the UAV's position, velocity, and time information by receiving satellite signals. This type of navigation system is known as a combined IMU/GNSS navigation system.


IMU/GNSS INS uses a 22-axis Extended Kalman Filter (EKF) structure to estimate pose in the NED reference frame. The state is defined as $\hat{X}_k=\{\hat{x}_1, \ldots, \hat{x}_n| n \in (1,22)  \}$, where the definition of each axis $\hat{x}_i$ is illustrated in Table~\ref{axis}.

\begin{table}[htb]
 \centering \caption{Element and Meaning of EKF Vector}
 \label{axis}
\begin{tabular}{c|c|c}
    \toprule[1.5pt]
    \textbf{Element} & \textbf{Label} & \textbf{Meaning}  \\
    \midrule[1pt] 
    
   $\hat{x}_1$& $\mathbf{q_0}$ &  \multirow{4}{*}{\makecell[c]{Orientation quaternion. }}\\\cline{1-2} 
   	
   $\hat{x}_2$&$\mathbf{q_1}$ & \\\cline{1-2} 
    
   $\hat{x}_3$& $\mathbf{q_2}$& \\\cline{1-2} 

   $\hat{x}_4$& $\mathbf{q_3}$&   \\\hline
   $\hat{x}_5$&$\mathbf{P_N}$ & \multirow{3}{*}{\makecell[c]{ UAV Position in local \\ NED coordinate system.} }  \\\cline{1-2} 
   $\hat{x}_6$&$\mathbf{P_E}$&   \\\cline{1-2} 
   $\hat{x}_7$&$\mathbf{P_D}$&   \\\hline
   $\hat{x}_8$& $\mathbf{V_N}$& \multirow{3}{*}{\makecell[c]{ UAV Velocity in local\\ NED coordinate system.}  }\\\cline{1-2}
   $\hat{x}_9$&$\mathbf{V_E}$ &   \\\cline{1-2}
   $\hat{x}_{10}$&$\mathbf{V_D}$ &   \\\hline
   $\hat{x}_{11}$& $\mathbf{\Delta \theta bias_X}$ &  \multirow{3}{*}{ \makecell[c]{Bias in integrated \\ gyroscope reading.}} \\\cline{1-2}
    $\hat{x}_{12}$& $\mathbf{\Delta \theta bias_Y}$ &   \\\cline{1-2}
    $\hat{x}_{13}$& $\mathbf{\Delta \theta bias_Z}$&   \\\hline
    $\hat{x}_{14}$& $\mathbf{\Delta v bias_X}$& \multirow{3}{*}{\makecell[c]{Bias in integrated \\ accelerometer reading.}}  \\\cline{1-2}
    $\hat{x}_{15}$& $\Delta v bias_Y$&   \\\cline{1-2}
    $\hat{x}_{16}$& $\mathbf{\Delta v bias_Z}$&   \\\hline
    $\hat{x}_{17}$& $\mathbf{geomagneticField_N}$ &  \multirow{3}{*}{ \makecell[c]{Estimate of geomagnetic \\field vector at the\\ reference location.}} \\\cline{1-2}
    $\hat{x}_{18}$& $\mathbf{geomagneticField_E}$&   \\\cline{1-2}
    $\hat{x}_{19}$& $\mathbf{geomagneticField_D}$&   \\\hline
    $\hat{x}_{20}$& $\mathbf{magbias_X}$ & \multirow{3}{*}{\makecell[c]{Bias in the \\magnetometer readings.}}  \\\cline{1-2}
    $\hat{x}_{21}$& $\mathbf{magbias_Y}$ &   \\\cline{1-2}
    $\hat{x}_{22}$& $\mathbf{magbias_Z}$ &   \\
   \bottomrule[1.5pt]
     \end{tabular}
\end{table}

Firstly, The system model $f(\cdot)$ uses the estimated previous state $\hat{X}_{k-1}$ and a control input $u_{k-1}$, to predict the current state $\hat{X}_k^-$.
\begin{equation}
\label{f}
\hat{X}_k^- = f(\hat{X}_{k-1}, u_k)
\end{equation}
where $u_k$ are control inputs, typically angular velocity and acceleration data from IMU. 

After prediction, INS predicts the measurement $z_k$ at time $k$ for updating the current state. we define as: 
\begin{equation}
\label{z}
    z_k=h(\hat{X}_k)+v_k
\end{equation}
where $z_k$ comprises magnetic field data from magnetometers, gravitational acceleration data from accelerometers, and position and velocity from GNSS. $h(\cdot)$ is the measurement prediction model and $v_k$ is the observational noise.

Due to hardware arithmetic limitations, INS often handles nonlinear functions by truncating their Taylor expansions with first-order linearization and neglecting the higher-order terms. This approach transforms the nonlinear problem into a linear one, as exemplified by Eq. \ref{nolinerf1} and \ref{nolinerf2}:
\begin{equation}
\label{nolinerf1}
f(X_{k-1}, u_k) \approx f(\hat{X}_{k-1}, u_k) + \frac{\partial f}{\partial x}\bigg|_{\hat{X}_{k-1}, u_k} (X_{k-1} - \hat{X}_{k-1})
\end{equation}

\begin{equation}
\label{nolinerf2}
h(X_k) \approx h(\hat{X}_k^-) + \frac{\partial h}{\partial x}\bigg|_{\hat{X}_k^-} (X_k - \hat{X}_k^-)
\end{equation}
INS defines the state transfer matrix $F_K$ and the Jacobi matrix of the measurement $H_k$ respectively:
\begin{equation}
    F_k = \frac{\partial f}{\partial x}\bigg|_{\hat{X}_{k-1}, u_k}
\end{equation}

\begin{equation}
     H_k = \frac{\partial h}{\partial x}\bigg|_{\hat{X}_k^-} 
\end{equation}

Based on Eq. \ref{f} and \ref{z}, we can derive the priori estimated covariance matrix  $P_k^-$ at time $k$.
\begin{equation}
\label{e}
    e_k=\hat{X}_k - \hat{X}_k^-
\end{equation}
\begin{equation}
\label{p}
\begin{split}
   P_k^- &=E(e_k e_k^T)\\
   & = F_k P_{k-1} F_k^T + Q_k
\end{split}
\end{equation}
where $Q_k$ refers to the process noise covariance matrix at time step $k$.

When obtaining the measurement $z_k$, the system will calculate the Kalman gain $K_k$ and update the estimated state to obtain an accurate estimation of the state information in the following way:
\begin{equation}
\label{K}
K_k = P_k^- H_k^T (H_k P_k^- H_k^T + R_k)^{-1}
\end{equation}

\begin{equation}
\hat{X}_k = \hat{X}_k^- + K_k(z_k - h(\hat{X}_k^-))
\end{equation}

\begin{equation}
P_k = (I - K_k H_k) P_k^-
\end{equation}
where $R_k$ refers to the measurement noise covariance matrix at time step $k$.

\subsection{Detector}
In dynamic system state estimation, the EKF optimizes the estimation of the system state successively through prediction and update steps. It computes the residual $r(k)=z_k - h(\hat{x}_k^-)$ in each step to reflect the difference between the actual measured value and the predicted value. With no attacks or anomalies, $r(k)$ will be presented as a zero-mean Gaussian distribution with a covariance matrix $Pr:=H_k P_k H_k^T+R_k$.

However, the system may generate outliers due to attacks, noise, faults, and other factors, all of which contribute to the measurements deviating from the true values. To prevent the EKF state from these disruptive outliers, implementing an outlier detection mechanism becomes crucial. 
The chi-square statistical test serves as an efficient tool for determining outliers~\cite{schreiber2016vehicle,piche2016online}. It evaluates the current measured value by calculating the chi-square statistic $\chi^2_k$, comparing it to a predefined statistical significance threshold. When $\chi^2_k$ surpasses this threshold $\tau$, the measurement is considered an outlier, and suitable measures are undertaken, including discarding the measurement or executing a partial update. The chi-square statistic $\chi^2_k$ is defined as:

\begin{equation}
\begin{aligned}
\chi^2_k = r(k)^T S_k r(k)
\\
S_k=(H_k P_k^- H_k^T + R_k)
\end{aligned}
\end{equation}

\section{Security Analysis}
\subsection{Attack Formulation}
Viewed from the perspective of navigation equations, the process of a spoofing attack on GNSS signals by an attacker can be described as follows: the attacker injects $n$ spoofing signal $\{\delta _k|k=1,\dots,n\}$ into the measurement data, resulting in a modification of the measurement $h$ as follows.
\begin{equation}
    z_k=h(\hat{X}_k)+\delta _k+v_k
\end{equation}

Due to the higher occurrence of data errors in the GNSS \textit{z-axis} and the availability of alternative altitude data sources, only the \textit{NE} (North-East) directional updates are applied to the position vector. As for the attacker, they can only modify the position and speed provided by GNSS, i.e., dimensions 5-6 and 8-10 in Table \ref{axis}.
%
%
\subsection{Study of Attack}
\label{sec:study}
The GNSS observation matrix $H_{GNSS}$ are as follows.
\begin{equation}
 H_{GNSS}= 
     \begin{bmatrix}
        0_{1\times4}&1 & 0& 0&0_{1\times3} & 0_{1\times14}\\ 
        0_{1\times4}& 0& 1& 0& 0_{1\times3} & 0_{1\times14} \\
        0_{1\times4}& 0 & 0 & 0&0_{1\times3} &0_{1\times14} \\
        0_{3\times4}&0_{3\times1} &0_{3\times1} & 0_{3\times1}&I_{3\times3} & 0_{3\times14} \\
     \end{bmatrix}
\end{equation}

Since $H_{GNSS}$ is a sparse matrix, when updating position and velocity, $K_k$ can be simplified to the following form:
\begin{equation}
\label{k1}
\begin{split}
    K_{k} &= P_k^- H_{GPS}^T (H_{GPS} P_k^- H_{GPS}^T + R_k)^{-1} \\
    &=(P_{k-1}+Q_k)(P_{k-1}+Q_k+R_k)^{-1} \\
    &=I-R_k(P_{k-1}+Q_k+R_k)^{-1}
\end{split}
\end{equation}

 From Eq.\ref{k1}, we can see that the state transfer error and the measurement error affect the magnitude of the gain $K_k$ simultaneously. \textbf{$Q_k$ and $R_k$ reflect the ability to cover systematic uncertainty and measurement uncertainty, respectively}. Therefore, inappropriate $Q_k$ and $R_k$ can lead to filter divergence or biased estimation. However, in INS, $Q_k$ and $R_k$ are generally determined based on a priori knowledge by pre-running the filter calculations offline and remain constant during the filtering process online. As a result, both the process estimation error covariance $R_k$ and the Kalman gain $K_k$ converge quickly and remain constant during the process, demonstrating that the value of $K_k$ is determined by the ratio of $Q_k$ and $R_k$.
 
We assume the adversarial adds an  $\delta_i$ at time $i$. The prediction equation for the EKF becomes
\begin{equation}
\begin{split}
    \hat{x}_i^a &= \hat{x}_i^- + K_i(z_i + \delta _i - h(\hat{x}_i^-, 0))\\
    &=\hat{x_i}+K_i \delta _i
\end{split}
\end{equation}

\begin{equation}
P_k = (I - K_k H_k) P_k^-
\end{equation}
Therefore, when the UAV performs maneuvers, the impact of spoofing on localization results can be described in the following two ways: 
\begin{itemize}
    \item \textbf{Q uncertainty:} EKF uses Euler integrals to update the positional status, i.e:
    \begin{equation}
       \begin{bmatrix}  
    P_{N} \\ P_{E} \\ P_{D}\\  
  \end{bmatrix}_{i+1} =
  \begin{bmatrix}  
    P_{N} \\ P_{E} \\ P_{D}\\  
  \end{bmatrix}_i+
  \begin{bmatrix}  
    V_{N} \\ V_{E} \\ V_{D}\\  
  \end{bmatrix}_i
  \Delta t
    \end{equation}
    Firstly, when performing maneuvers, the system is highly nonlinear. i.e.,
    \begin{equation}
     \exists \epsilon >0, \|\frac{d\mathbf{v_i}}{dt}\| \geq \epsilon 
    \end{equation}
    The acceleration $\mathbf{a_i \neq 0}$ and the update equation for position essentially becomes:
        \begin{equation}
       \begin{bmatrix}  
    P_{N} \\ P_{E} \\ P_{D}\\  
  \end{bmatrix}_{i+1} =
  \begin{bmatrix}  
    P_{N} \\ P_{E} \\ P_{D}\\  
  \end{bmatrix}_k +
  \begin{bmatrix}  
    V_{N} \\ V_{E} \\ V_{D}\\  
  \end{bmatrix}_i
  \Delta t+
   \begin{bmatrix}
      \frac{1}{2}\Delta t^2,0,0 \\ 0,\frac{1}{2}\Delta t^2,0 \\ 0,0,\frac{1}{2}\Delta t^2\\
  \end{bmatrix}
  \begin{bmatrix}
      a_{N} \\ a_{E} \\ a_{D}\\
  \end{bmatrix}
  \end{equation}
    According to Eq. \ref{e} and \ref{p}, the accumulation of linearisation errors will increase $e_i$ and thus increase the process noise $P_i$. leading to inaccurate mathematical modeling and huge nonlinear errors. The fixed $Q_k$ makes it difficult to suppress the nonlinear error increased due to the change of motion state. According to Eq. \ref{k1}, the value of $K_i$ will indirectly increase, making the INS more inclined to trust the GNSS data.  In addition, this complex nonlinear characteristic will be further expanded due to physical factors such as the lag of IMU data and the presence of friction in the gyroscope. Therefore, in this scenario, the prediction of the system model cannot effectively reflect the actual physical process.
    
    \item \textbf{R uncertainty:} The modification of GNSS results in a shift in the measurement data distribution, rendering a fixed $R_k$ inadequate for accurately describing the measurement noise distribution. Consequently, $\delta_i$ experiences a significant increase. Additionally, the rising $K_k$ value leads the system to place greater trust in GNSS, further amplifying the impact of the attack associated with $\delta_i$.
\end{itemize}

\begin{figure*}[htbp]
	\centering
	\begin{subfigure}{0.24\linewidth}
		\centering
        		\includegraphics[width=0.9\linewidth]{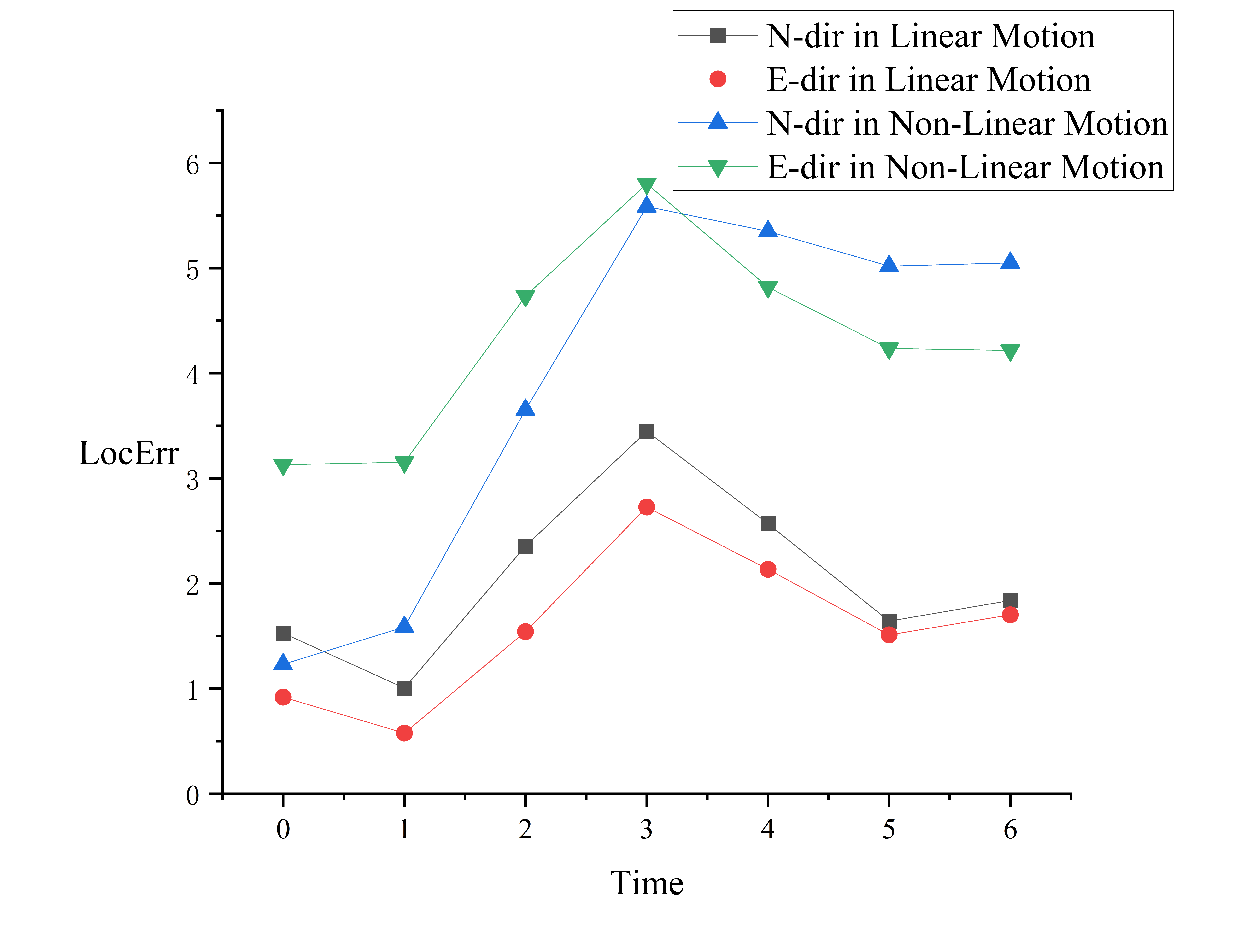}
		\caption{Bias Signal Attack}
		\label{fig:finding11}
	\end{subfigure}
	\centering
	\begin{subfigure}{0.24\linewidth}
		\centering
        \includegraphics[width=0.9\linewidth]{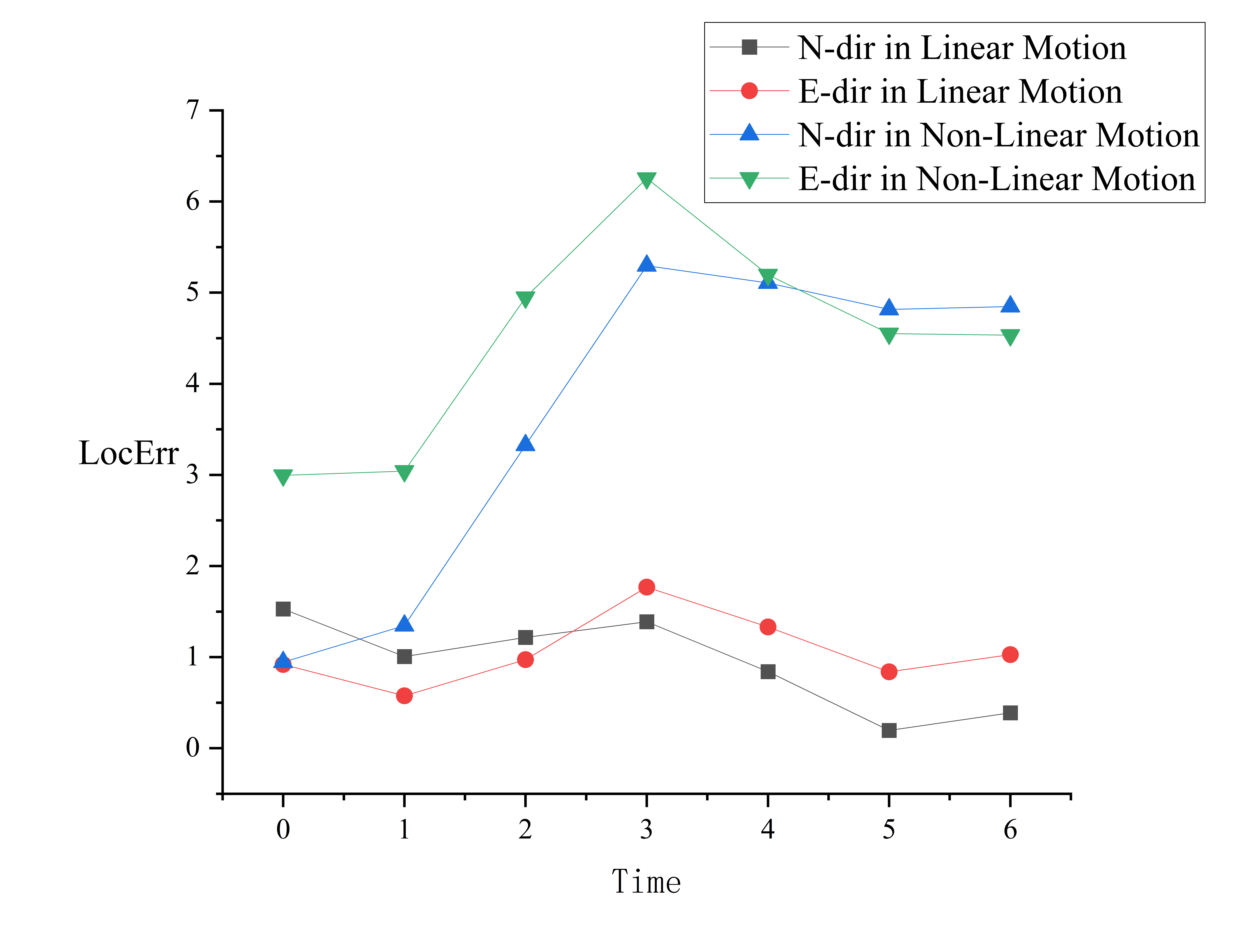}
		\caption{Multiplicative Attack}
		\label{fig:finding12}
	\end{subfigure}
    	\centering
	\begin{subfigure}{0.24\linewidth}
		\centering
        \includegraphics[width=0.9\linewidth]{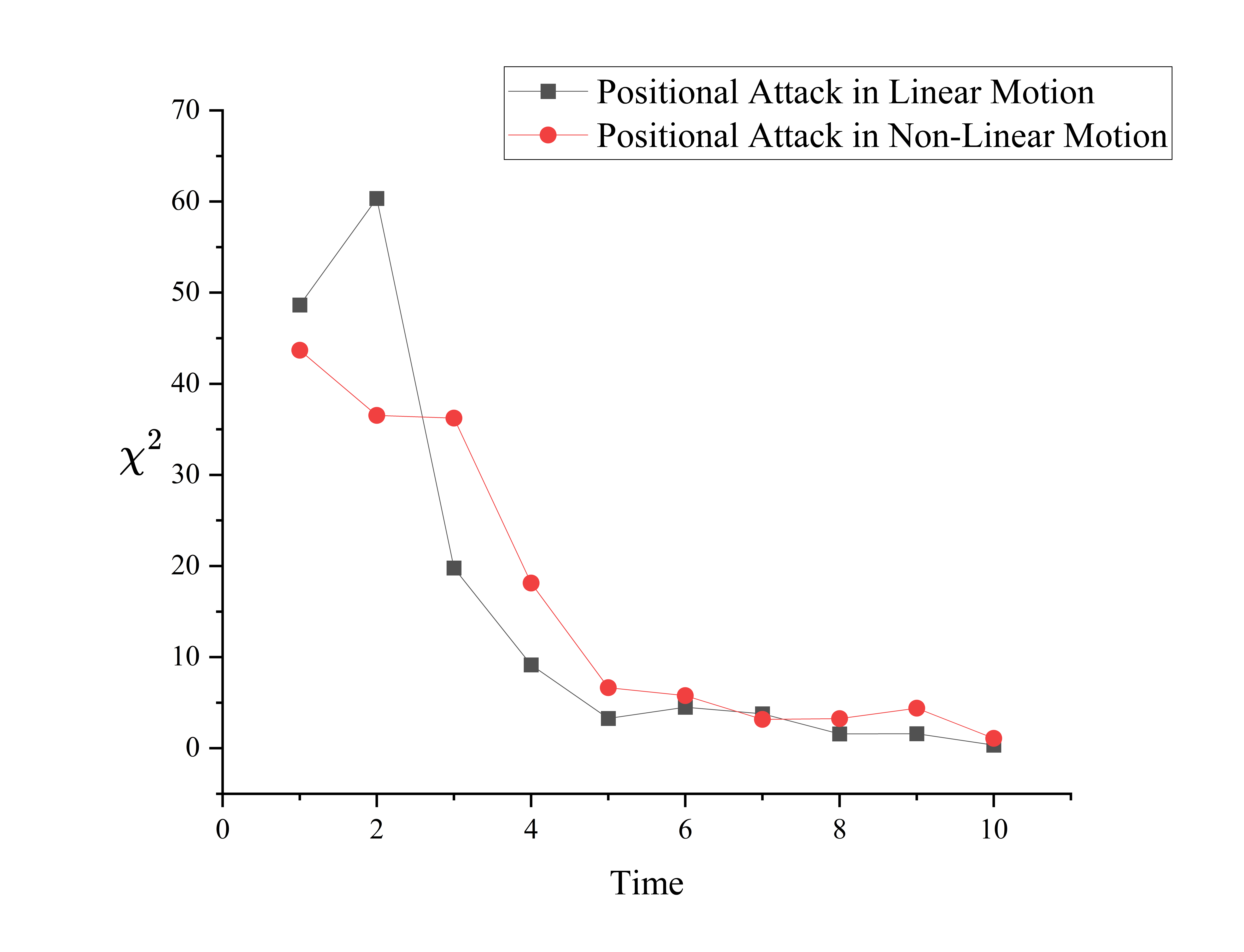}
		\caption{Positional Attack}
		\label{fig:study21}
	\end{subfigure}
	\centering
	\begin{subfigure}{0.24\linewidth}
		\centering
        \includegraphics[width=0.9\linewidth]{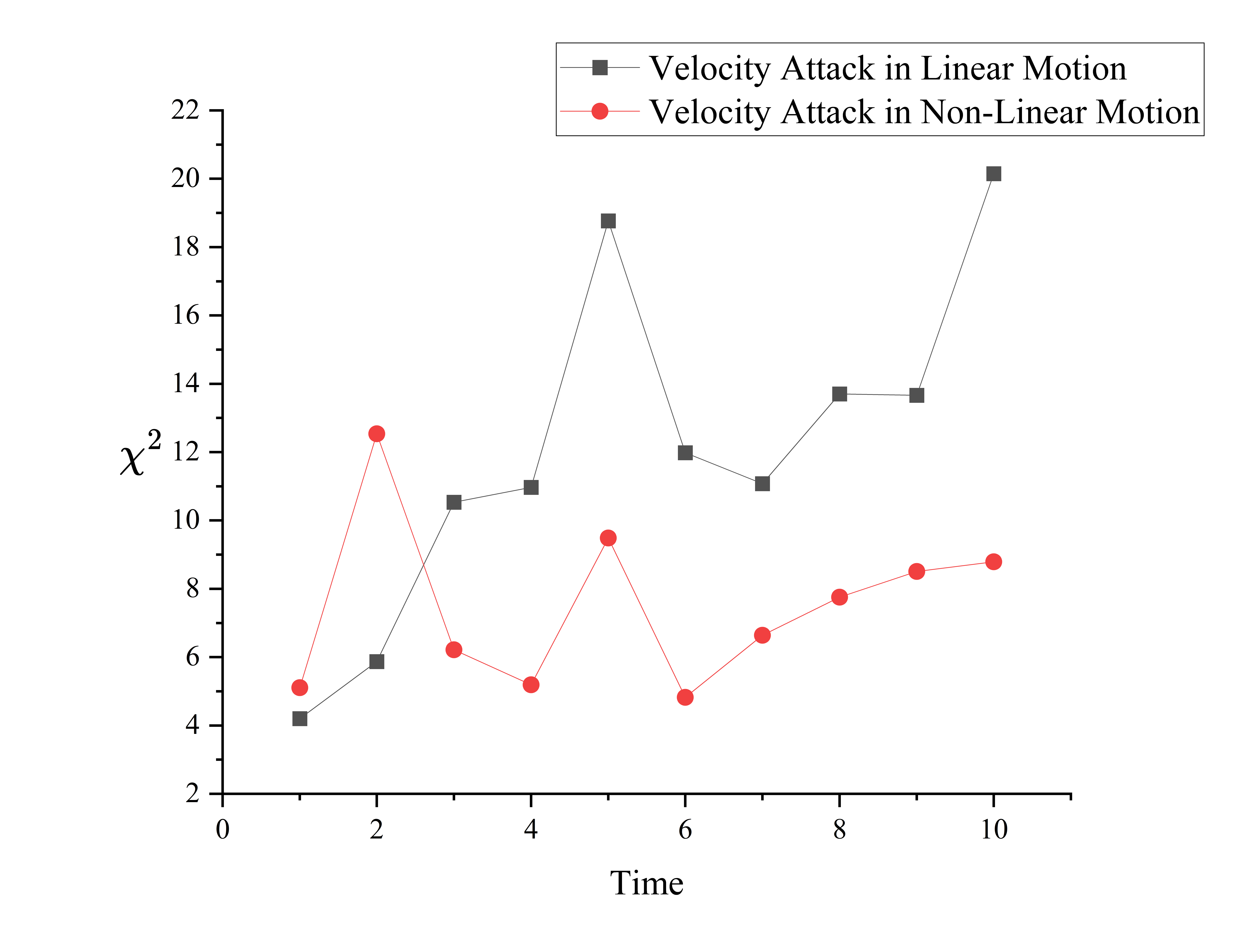}
		\caption{Velocity Attack}
		\label{fig:study22}
	\end{subfigure}
 \caption{Study of GNSS Attack under Different Motion states}
	\label{fig:study}
\end{figure*}

We employ a biased signal attack and a multiplicative attack to evaluate the phenomenon above. The UAV maintains a constant velocity of $5m/s$ and performs two typical motion modes, uniform linear motion (linear motion) and uniform circular motion (non-linear motion), respectively. For each flight state, we apply an attack window of two attack inputs for the GNSS respectively and observe the changes in the localization Error $LocErr$ before and after the attack. As a result, the attack time is 2 seconds since the GNSS input is 1 Hz. The experimental results are shown in Fig. \ref{fig:finding11} and \ref{fig:finding12}. Thus, we can get Finding 1.

\begin{tcolorbox}[left=1mm, right=1mm, top=0.5mm, bottom=0.5mm, arc=1mm]
\textbf{Finding 1:} \textit{For a GNSS spoofing of the same magnitude, applying it during the UAV's non-linear motion results in more pronounced fluctuations in positioning accuracy compared to linear motion. These changes in motion dynamics create greater vulnerabilities for INS.}
\end{tcolorbox}

Position and velocity are tightly coupled in INS, making velocity attacks easier to modify the position result.  However, there has been limited research on attacks targeting velocity. We would like to explore one question: Are there obvious
correlations between the velocity attack stealthiness and the UAV motion states? To validate this question, we select two flight trajectories with linear and nonlinear motion states respectively. We use a biased signal attack to perturb velocity and position measurement and evaluate stealthiness by observing the changes in the chi-square detector. Fig. \ref{fig:study21} and \ref{fig:study22} indicate that the detector shows no significant fluctuations between the two motion states for the positional attack. However, velocity attacks demonstrate a greater sensitivity to motion states, with significantly lower cardinality detector results observed in nonlinear motion states. This reveals key insights, summarized as Finding 2.

\begin{tcolorbox}[left=1mm, right=1mm, top=0.5mm, bottom=0.5mm, arc=1mm]
\textbf{Finding 2:} \textit{
The nonlinear motion state does not significantly affect the stealthiness of positional attacks, but it enhances the stealthiness of velocity attacks.}
\end{tcolorbox}

\section{Attack Design}
From the analysis of Section \ref{sec:study}, we observe that UAV exhibits greater vulnerability in a non-linear motion state compared to a linear one due to the combined effects of model and measurement uncertainty. During non-linear motion, an attack of the same magnitude can produce a more significant change in the navigation output than in the linear motion state. However, this vulnerability arises only when the uncertainty is heightened due to changes in the motion state. Moreover, \textbf{Finding2} indicates that a positional attack in non-linear motion will increase the risk of being detected. This introduces a key challenge to the attacker:  

\textbf{C1: How to opportunistically exploit these vulnerable periods to achieve maximum localization error while maintaining stealth.}

To address \textbf{C1}, we design a backdoor-like attack to exploit these vulnerable periods directly.  Inspired by the backdoor attacks in deep networks (DNNs)~\cite{chen2022clean,han2022physical,han2024backdooring}, we proposed a novel stated-based stealthy backdoor (SSD) attack against INS in a route planning scenario. SSD utilizes the motion state as a trigger, enabling the UAV to trigger an attack in a nonlinear motion state while maintaining normal operation in a linear motion state. Based on this design, not only can the detection rates of the attack be greatly reduced, but also the mission completion rate of the UAV can be effectively reduced. (UAV mission completion is usually accompanied by large maneuvers.)

\begin{equation}
\label{line}
\exists c \in \mathbb{R}, \|\frac{d\mathbf{v_i}}{dt}\| \leq c \quad \forall t \in [t_0, t_1]
\end{equation}
Specifically, when the victim UAV is in a stable linear flight state (Equation~\ref{line}), SSD adds a bias $F(t_i;\theta,\alpha)$ to attack the position stealthily. 
\begin{equation}
     F(t_i;\theta,\alpha)=\theta {e^{t_i/\alpha}}
\end{equation}
When it is maneuvering, each dimension of the UAV's movement can experience both positive and negative acceleration, indicating acceleration and deceleration in that particular direction. We define the acceleration direction $\Vec{a}$ as:
\begin{equation}
\Vec{a}^d_i = \Vec{\frac{\partial v_i}{\partial t_i}}
\end{equation}
When the UAV undergoes acceleration in this dimension (i.e. $\Vec{a}^d_i$ \textgreater $0$ ), SSD makes smooth changes to the velocity by multiplying a stealthy velocity bias $G(t_i,a^d_i;\phi)$ to perturb the localization result. 
\begin{equation}
     G(t_i,a^d_i;\phi)=\log_{2}{(2+\phi \Vec{a}^d_i t_i)}
\end{equation}
Overall, SSD can be formalized as follows:
\begin{equation}
\begin{cases}
       X_i=X_i+ F(t_i;\theta,\alpha)
       \\
       V_i=V_i* G(t_i,a^d_i;\phi)
\end{cases}
\end{equation}
where $t_i$ is the attack time for attackers. $\theta$, $\alpha$, and $\phi$ are hyperparameters that can be dynamically adjusted to maintain an equilibrium between stealthiness and effectiveness. We will demonstrate how to configure these parameters in Section~\ref{pa}. SSD chooses acceleration as a trigger. It uses organic coupling between velocity and position to perform a combined attack. Its pseudocode is presented in Algorithm \ref{SSDa}.

\begin{algorithm}
    \caption{SSD}
    \label{SSDa}
    \LinesNumbered 
    \KwIn{Victim UAV position $P_i=(P_{Ni}, P_{Ei}, P_{Di})$, Victim UAV velocity  $V_i=(V_{Ni}, V_{Ei}, V_{Di})$, IMU sampling frequency $F_{imu}$, Iteration number $M$}
    Set initialize hyperparameters $\theta$,$\phi$ and $\alpha$ \;
    \For{i=1 to M}{
    \For{j=1 to $F_{imu}$}{
       Receive IMU data \;
       Predict $X_{i+1}$ using IMU data;  
    }
     Receive victim UAV velocity  $V_i$ and position $P_i$ from GNSS  \;
        Compute acceleration $a_i=(\frac{\partial V_{Ni}}{\partial t_i}, \frac{\partial V_{Ei}}{\partial t_i}, \frac{\partial V_{Di}}{\partial t_i})$\;
             \If{$||a_i||_2$=0}{
              // \emph{Apply position perturbation in linear motion state} \;
                 $P_i=P_i+ \theta {e^{ti/\alpha}}$\;

            }
        \Else{
        // \emph{Apply velocity perturbation in nonlinear motion state} \;
        $ V_{Ni}=V_{Ni}* \log_{2}{(2+{\frac{\partial v_{Ni}}{\partial t}}\phi t)}$\;
        $V_{Ei}=V_{Ni}* \log_{2}{(2+{\frac{\partial v_{Ei}}{\partial t}}\phi t)}$\;
        }
    Fuse $P_i$ and $V_i$ to update $X_{i+1}$
    }
    
\end{algorithm}

\begin{figure*}[htp]
	\centering
 \centerline{\includegraphics[width=380pt]{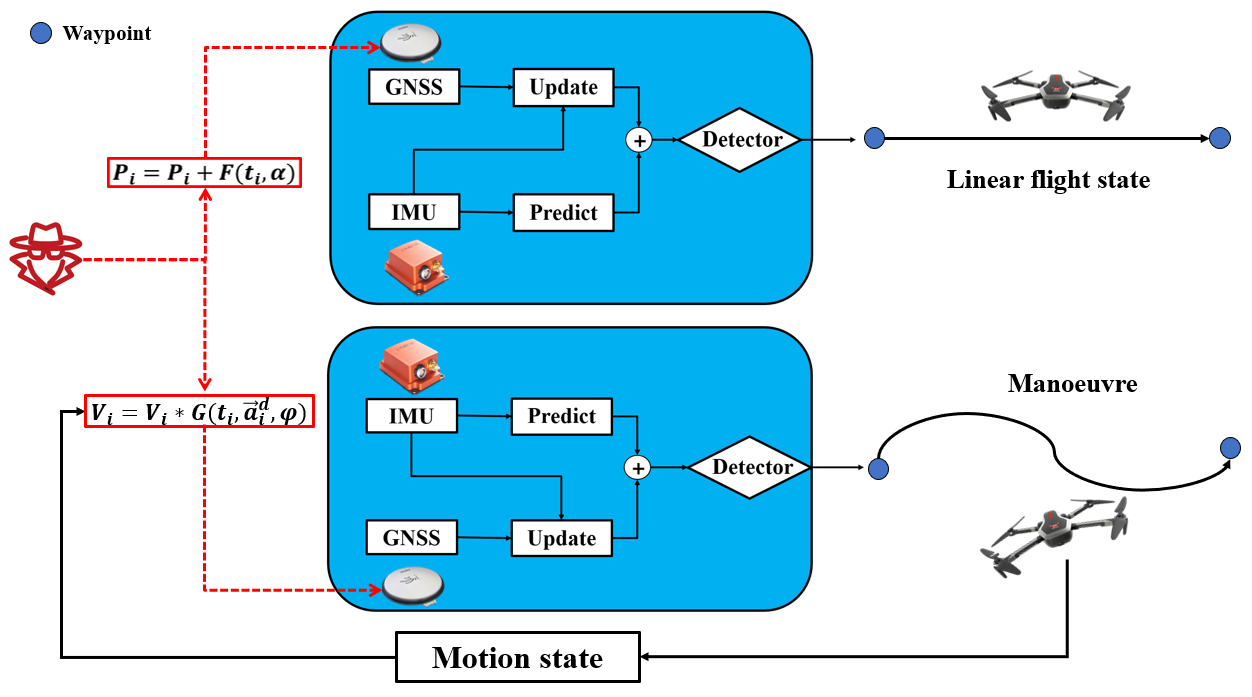}} 
		\caption{Overview of SSD}
		\label{fig:ssd}
\end{figure*}

\section{Experiment}
\subsection{Experimental Setup}
\subsubsection{Trajectory Dataset}
 We used three representative task trajectories. Trajectories \uppercase\expandafter{\romannumeral1} and \uppercase\expandafter{\romannumeral2} represent typical linear and non-linear motion states, respectively, while trajectory \uppercase\expandafter{\romannumeral3} combines both motion states.

\begin{figure*}[htbp]
	\centering
	\begin{subfigure}{0.325\linewidth}
		\centering
		\includegraphics[width=0.9\linewidth]{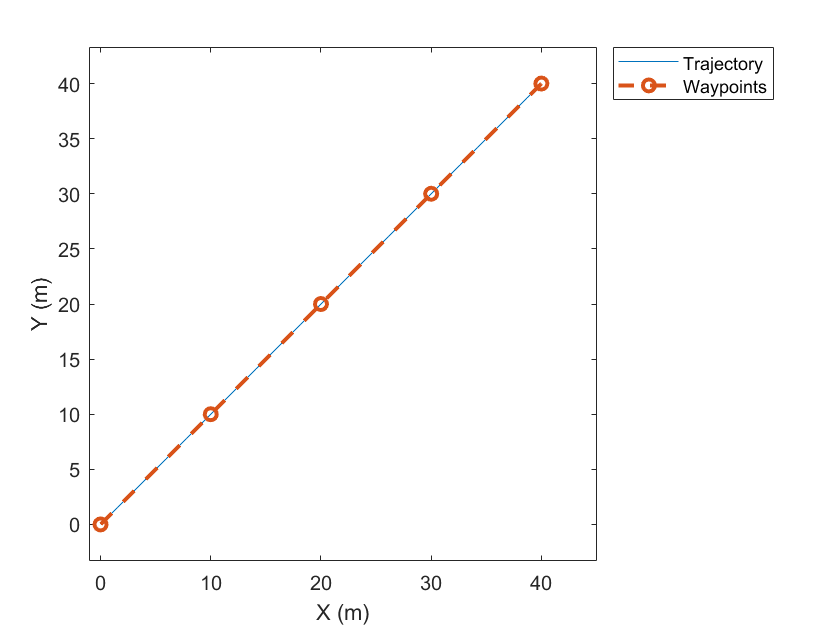}
		\caption{ Straight-line Path}
		\label{fig:t1}
	\end{subfigure}
	\centering
	\begin{subfigure}{0.325\linewidth}
		\centering
		\includegraphics[width=0.9\linewidth]{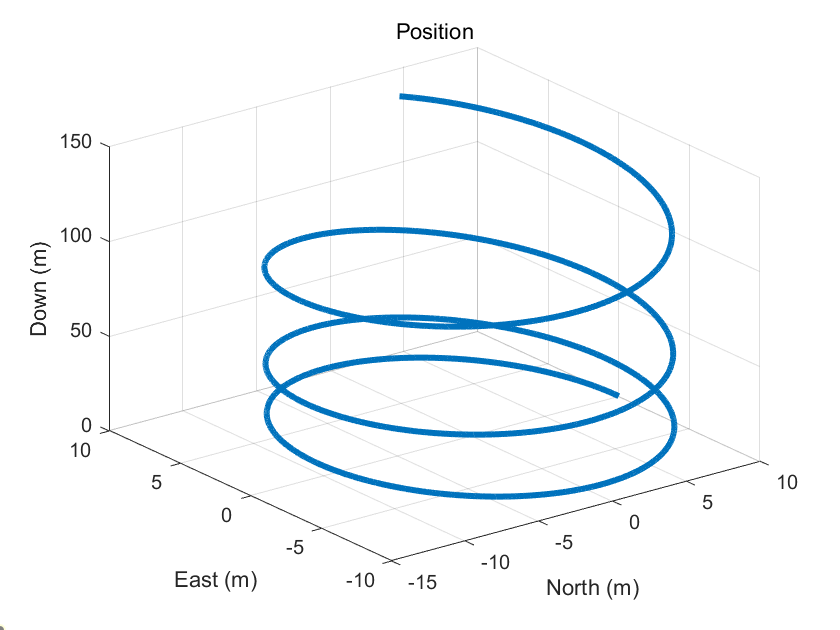}
		\caption{Spiral Path}
		\label{fig:t2}
	\end{subfigure}
	\centering
	\begin{subfigure}{0.325\linewidth}
		\centering
		\includegraphics[width=0.9\linewidth]{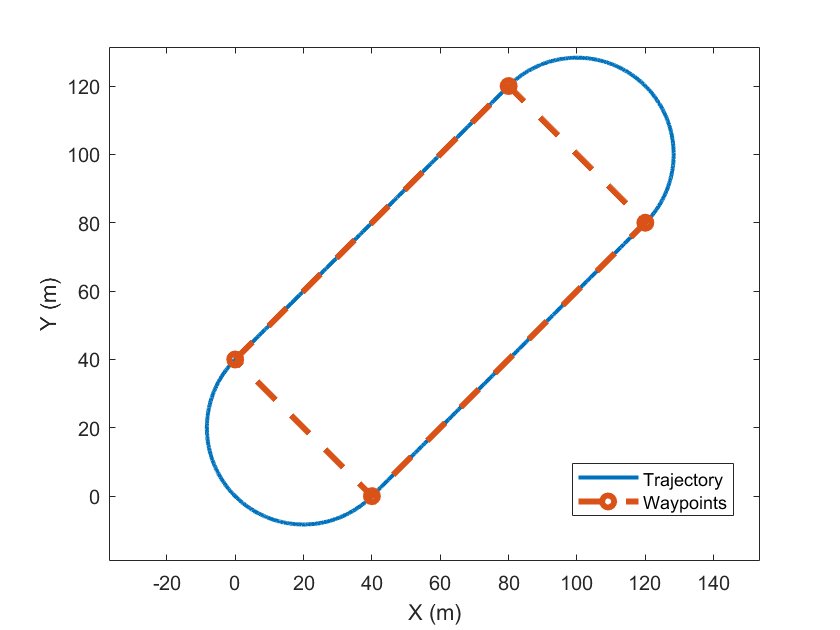}
		\caption{U-shape Path}
		\label{fig:t3}
	\end{subfigure}
	\caption{Trajectory Visualization}
	\label{fig:5}
\end{figure*}

\begin{itemize}
    \item \textbf{Trajectory \uppercase\expandafter{\romannumeral1}: Straight-line Path } As illustrated in Fig.~\ref{fig:t1}, this trajectory is designed for executing simple tasks, such as flying to a specific location and performing actions along a predefined path (e.g., patrols, cargo transport, etc.).
    
    \item \textbf{Trajectory \uppercase\expandafter{\romannumeral2}: Spiral Path:
} As illustrated in Fig.~\ref{fig:t2}, this trajectory is designed for tasks that involve changes in flight altitude, such as agricultural spraying, area scanning, and 3D mapping.

     \item \textbf{Trajectory \uppercase\expandafter{\romannumeral3}: U-shape Path:
} As illustrated in Fig.~\ref{fig:t3}, the trajectory is designed for tasks that demand high precision, such as monitoring, surveying, or round-trip transportation.


\end{itemize}

\subsubsection{Navigation Algorithm}
 We choose the estimation and control library EKF (ECL EKF2) of the PX4 drone autopilot \cite{eclekf2} project as the target navigation algorithms.
\begin{itemize}
\item \textbf{ECL EKF2} implements EKF to estimate pose in the NED reference frame by fusing MARG (magnetic, angular rate, gravity) and GNSS data. MARG data is derived from magnetometer, gyroscope, and accelerometer sensors. It uses a 22-element state vector to track the orientation quaternion, velocity, position, MARG sensor biases, and geomagnetic.

 \item \textbf{CD-EKF} is a variant of ECL EKF2. It implements a continuous-discrete EKF to estimate pose in the NED reference frame by fusing MARG and GNSS data. It uses a 28-element state vector to track the orientation quaternion, velocity, position, MARG sensor biases, and geomagnetic vector.
\end{itemize}

\subsubsection{Implementation details.}
\begin{figure}[htp]
	\centering
 \centerline{\includegraphics[width=230pt]{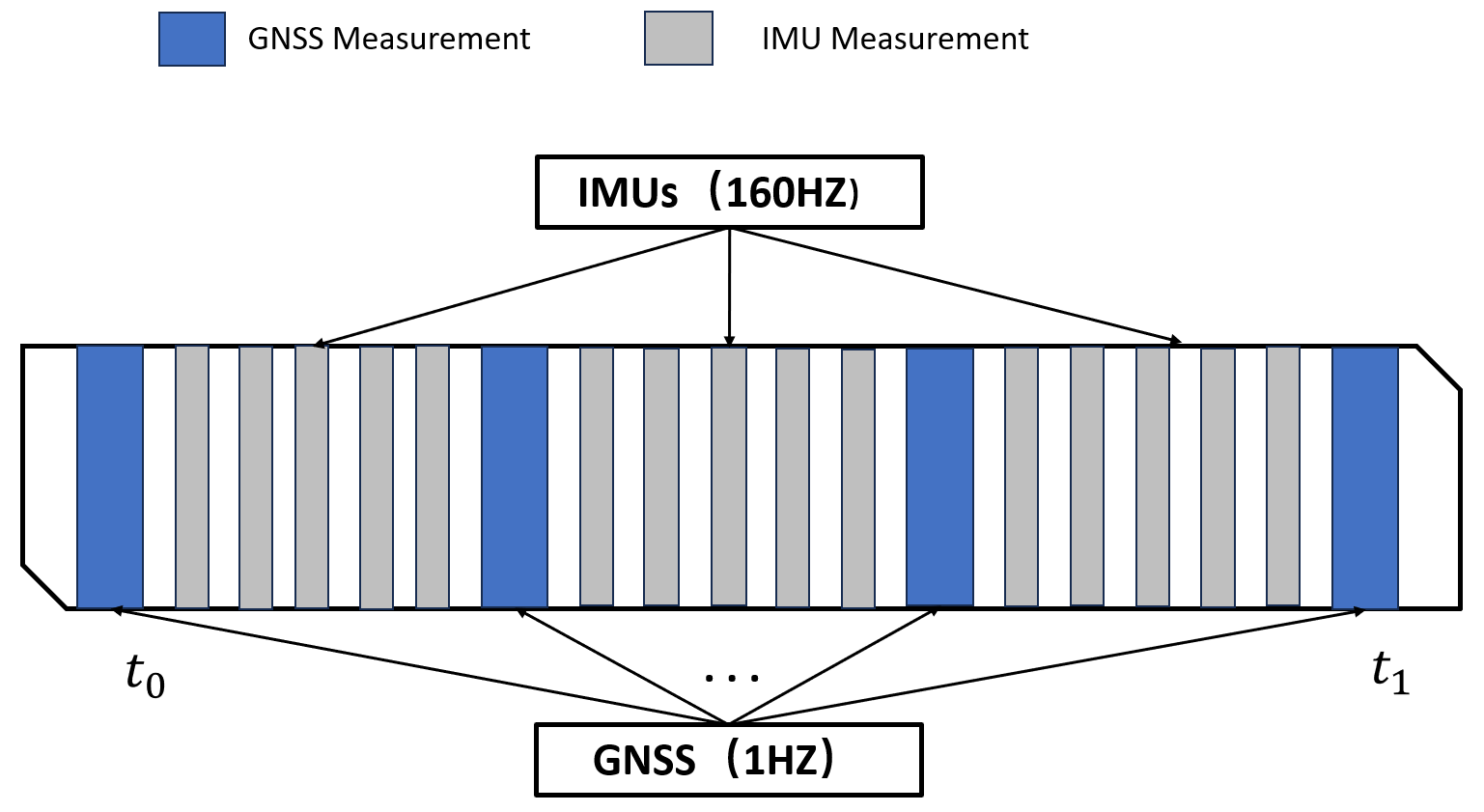}} 
		\caption{Modeling of IMU and GNSS Fusion}
		\label{fig:imu}
\end{figure}
Accelerometers and gyroscopes operate at relatively high sample rates and necessitate high-rate processing. In contrast, GNSS and magnetometers function at relatively low sampling rates and require lower data processing rates. To replicate this configuration in our experiment, the IMUs (accelerometers, gyroscopes, and magnetometers) were sampled at 160 Hz, while the GNSS was sampled at 1 Hz. As demonstrated in Fig~\ref{fig:imu}, only one out of every 160 samples from the IMUs was provided to the fusion algorithm.

\subsubsection{Evaluation Metircs}

We utilize three metrics to evaluate the effectiveness of SSD comprehensively. Initially, we include two metrics that are widely used in relevant studies \cite{zhang2022adversarial}.

(1) Average Displacement Error (\textbf{ADE}): This metric measures the average deviation between the predicted and ground-truth trajectories by calculating the root mean squared error (RMSE) across all time frames. It captures the overall accuracy of the predicted trajectory compared to the actual path.

(2) Final Displacement Error (\textbf{FDE}): FDE focuses specifically on the prediction accuracy at the final time frame, quantified as the RMSE between the predicted and ground-truth positions. This metric highlights the importance of precise final positioning, which is crucial in applications such as path planning.

However, the two indicators above alone are insufficient to capture the impact of targeted attacks on UAVs. This introduces another challenge for evaluating SSD:  

\textbf{C2: How to evaluate SSD's impact on UAV mission.}

In mission-critical scenarios, UAVs must execute precise manoeuvres at specific waypoints to ensure the successful completion of the mission. The deviation from these waypoints during flight significantly increases the likelihood of mission failure. Therefore, to address \textbf{C2}, we design the Average Per-Waypoint Displacement Error (\textbf{APDE}). The APDE is formally defined as the average of the displacement errors computed at each waypoint along the trajectory. We define APDE as follows:
\begin{equation}
\label{apde}
APDE=\frac{\sum_{i=1}^{N_w} \quad ||P_i-P_i^a||_2}{N_w}   
\end{equation}
where $P_i^a$ and $P_i$ refer to the predicted positions before and after the attack, respectively. $N_w$ refers to the number of waypoints. APDE provides a more nuanced understanding of how targeted attacks impact the UAV's ability to adhere to its prescribed flight path, particularly at critical waypoints, thus enabling a more accurate assessment of the potential risks to mission success.

\subsection{Parametric Analysis}
\label{pa}
\begin{figure*}[htbp]
	\centering
	\begin{subfigure}{0.49\linewidth}
		\centering
		\includegraphics[width=0.8\linewidth]{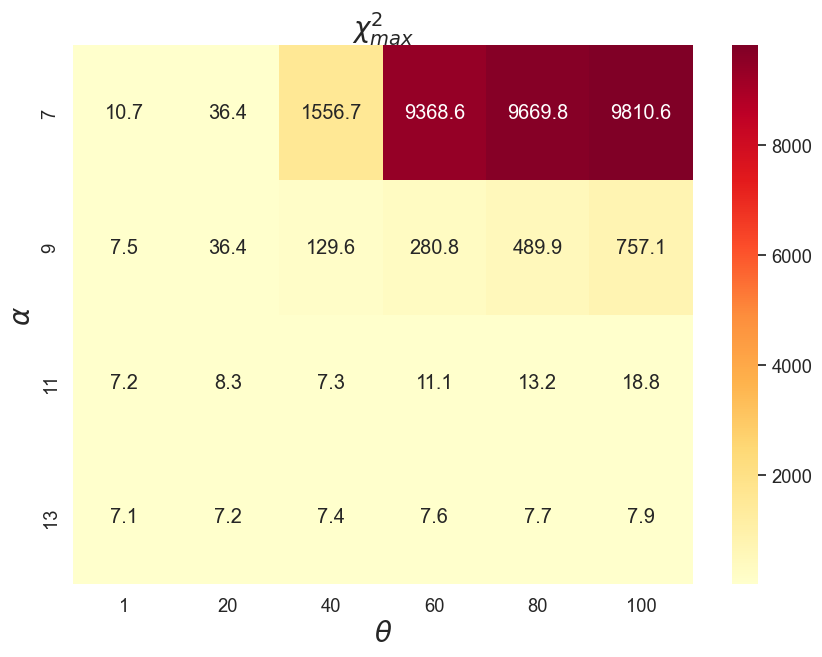}
		\caption{Positional attack}
		\label{fig:pa-a}
	\end{subfigure}
	\centering
	\begin{subfigure}{0.49\linewidth}
		\centering
		\includegraphics[width=0.8\linewidth]{./figure/FIG7B.png}
		\caption{Velocity  attack}
		\label{fig:pa-b}
	\end{subfigure}
 \caption{Parameter Analysis}
	\label{fig:pa}
\end{figure*}

The effectiveness and stealthiness of the attack are highly dependent on the choice of parameters. 
We analyze the sensitivity of SSD to various parameters by combining different values for $\theta$, $\alpha$, and $\phi$ across multiple scenarios (the same as Section~\ref{sec:study}). To measure the attack effectiveness and stealthiness, we use ADE and the maximum chi-square statistic (denoted as $\chi^2_{max}$), respectively. The experimental results are shown in Fig~\ref{fig:pa}. In positional attacks, as $\theta$ increases, $\chi^2_{max}$ exhibits a corresponding upward trend. We further observe that the rate of increase in $\chi^2_{max}$ slows as $\alpha$ increases. This suggests that higher values of $\alpha$ help mitigate the growth of $\chi^2_{max}$. Notably, when $\alpha$ exceeds a critical threshold value 13, $\chi^2_{max}$ starts to stabilize, converging within a relatively narrow threshold range. Within this range, $\chi^2_{max}$ fluctuates minimally and remains largely unaffected by changes in $\theta$, demonstrating high stability and consistency. For the velocity attack, as shown in Fig.~\ref{fig:pa-b}, $\chi^2_{max}$ gradually converges to about 6 when $\phi$ is less than 0.08. In the following experiments, we set the values of $\theta$, $\alpha$, and $\phi$ to 20, 11, and 0.08, respectively.

\subsection{Ablation study}
To validate the effectiveness of velocity-based and position-based attacks, we conducted an experiment where the UAV performed a 35 second flight incorporating linear and nonlinear motion states. The first 20 seconds involved uniform linear motion at a velocity of $2m/s$. Afterward, the UAV transitioned into a uniform circular motion mode, maintaining a linear velocity of $2m/s$ for the remaining 15 seconds. This flight trajectory was used for the ablation experiments.

\subsubsection{Contributions of different attacks} 
\begin{figure}[htp]
	\centering
 \centerline{\includegraphics[width=230pt]{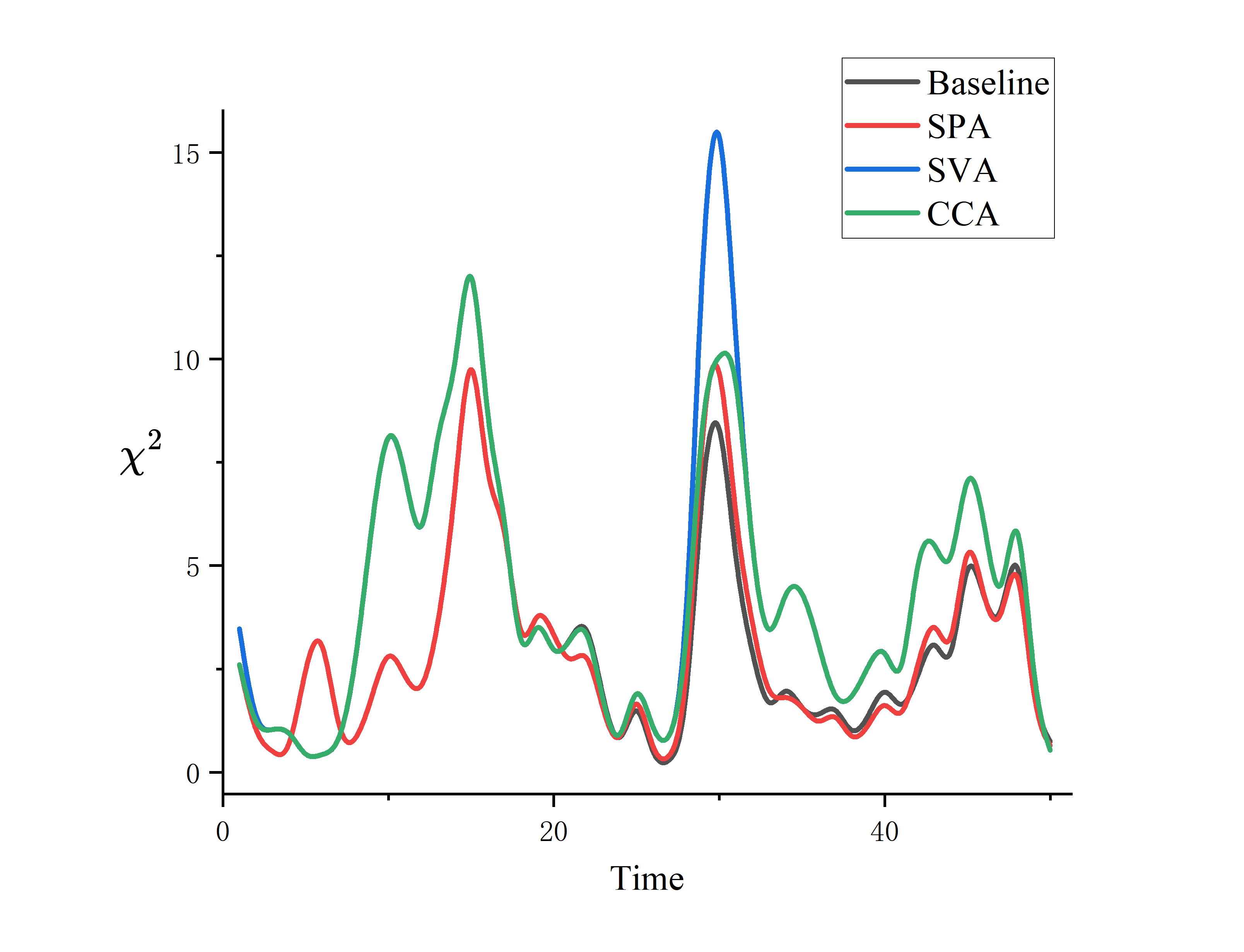}} 
		\caption{Stealthiness Comparison for Attack Contribution}
		\label{fig:ablation}
\end{figure}

To explore the contribution of velocity and positional attacks to overall attack effectiveness, we designed comparison experiments with three attack strategies: single position perturbation attack (SPA), single velocity perturbation attack (SVA), and combined concerted attack (CCA). The quantitative analysis of ADEs, presented in Table~\ref{ablation}, shows that both individual attacks are effective. Specifically, the ADEs for SPA and SVA are improved by 184\% and 212\%, respectively, compared to the baseline. When comparing the combined attack (CCA) to the single attacks, the ADE improves by 129\% over SPA and 112\% over SVA. This indicates that CCA not only significantly enhances attack effectiveness by leveraging the coupling effect between position and velocity but also perturbs both longitude and latitude directions simultaneously. Furthermore, the stealth assessment results in Fig.~\ref{fig:ablation} show that while the CCA approach induces brief oscillations in the detection statistics, these fluctuations remain well within the acceptable threshold limits, ensuring that the attack remains stealthy.
\begin{table}[htb]
 \centering \caption{Ablation Study of Different Attacks' Contribution}
 \label{ablation}
\begin{tabular}{ccc}
    \toprule[1.5pt]
      & \textbf{Position N (meters)} & \textbf{Position E (meters)} \\
    \midrule[1pt]
    
    \textbf{Baseline} & 1.41& 0.66 \\\hline
    \textbf{SPA} & 2.72 & 0.89\\\hline
    
    \textbf{SVA}& 2.45 & 2.2  \\\hline

    \textbf{CCA}& 2.79 & 2.41 \\
    \bottomrule[1.5pt]
     \end{tabular}
\end{table}

%

\subsubsection{Attack Combination}
\begin{figure}[htp]
	\centering
 \centerline{\includegraphics[width=230pt]{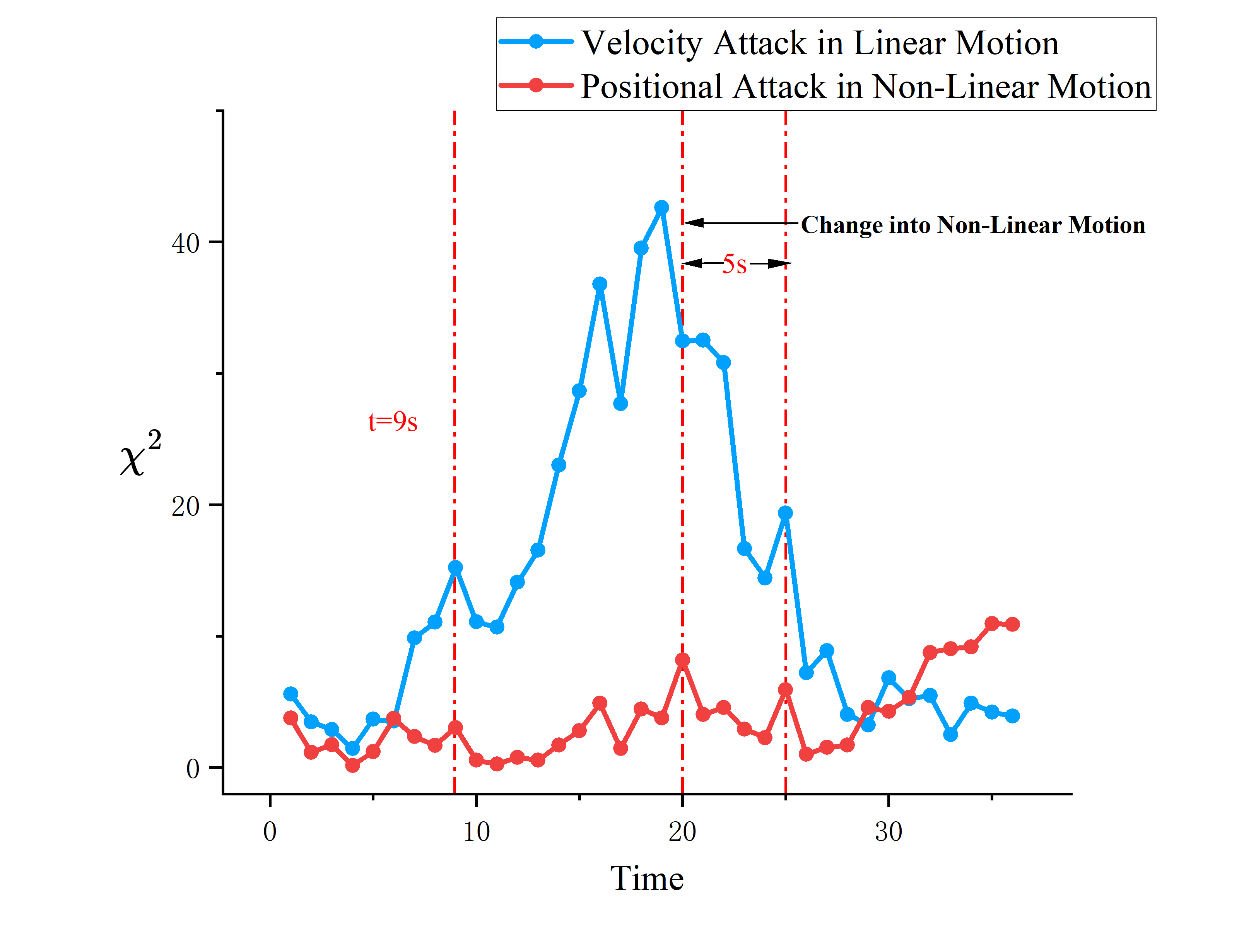}} 
		\caption{Detection Statistic Comparison for Attack Combination }
		\label{fig:ablation1}
\end{figure}

We apply the velocity attack during the linear motion state and the position attack during the nonlinear motion state and examine the impact of this combination on the attack's stealthiness. The experimental results, shown in Fig.~\ref{fig:ablation1}, reveal that $9s$ after the attack begins, the detector quickly identifies the existence of the attack. Furthermore, even when the UAV transitions to a nonlinear motion state, the detector continues to successfully detect the attack for $5s$, despite the change in attack mode. This phenomenon occurs because the drastic velocity changes in the linear motion state cause significant fluctuations in the residuals. In contrast, the nonlinear motion state allows SSD to effectively smooth out the impact of the velocity attack, ensuring the attack remains stealthy throughout the process. This result, along with \textbf{Finding2}, further validates the rationale of the SSD framework.

\subsection{Attack Stealthiness}
\label{section:stealthy}
\begin{figure*}[htbp]
	\centering
	\begin{subfigure}{0.325\linewidth}
		\centering
		\includegraphics[width=0.9\linewidth]{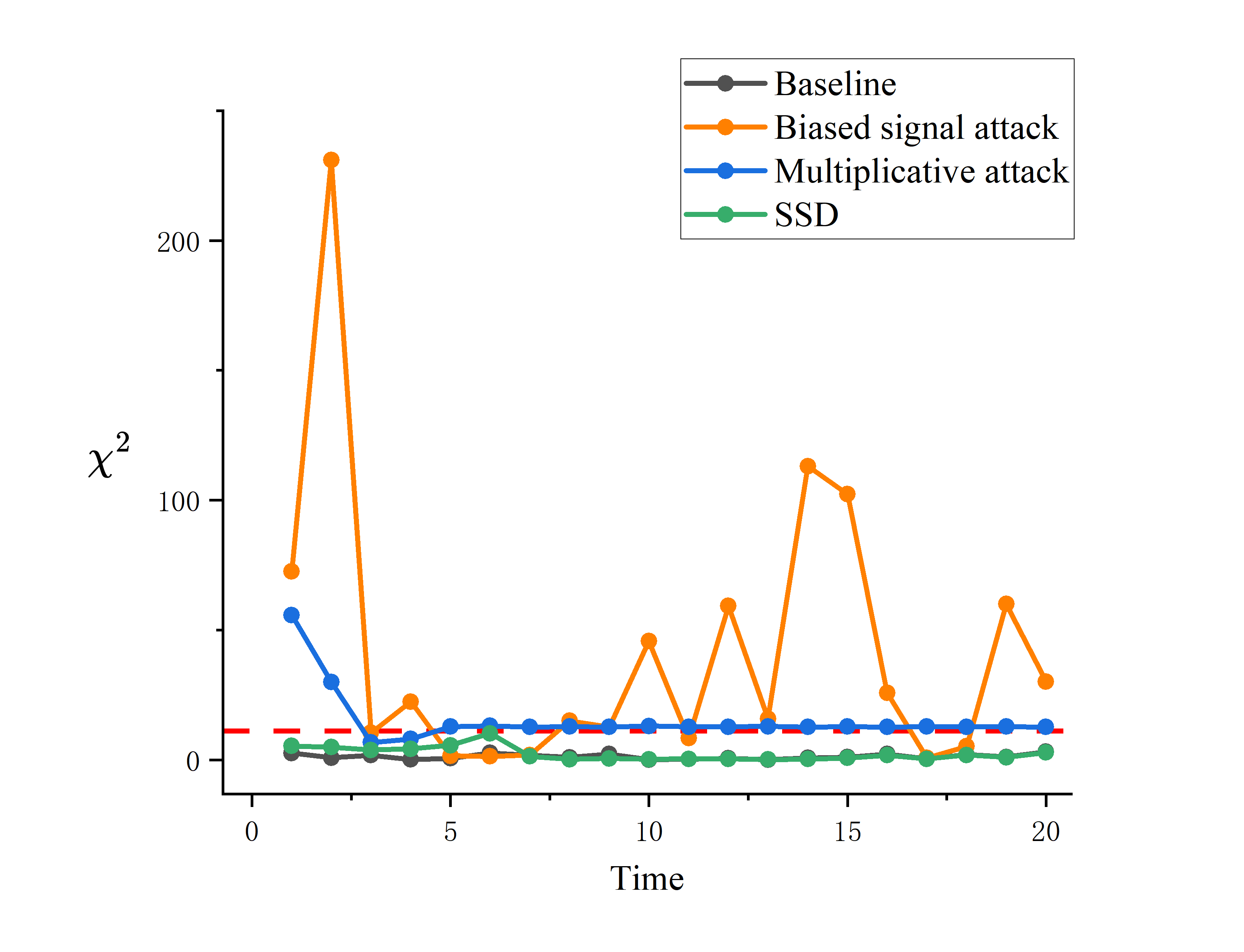}
		\caption{ $\chi^2 $ detection of different attacks under Trajectory \uppercase\expandafter{\romannumeral1} }
		\label{fig:ste1}
	\end{subfigure}
	\centering
	\begin{subfigure}{0.325\linewidth}
		\centering
		\includegraphics[width=0.9\linewidth]{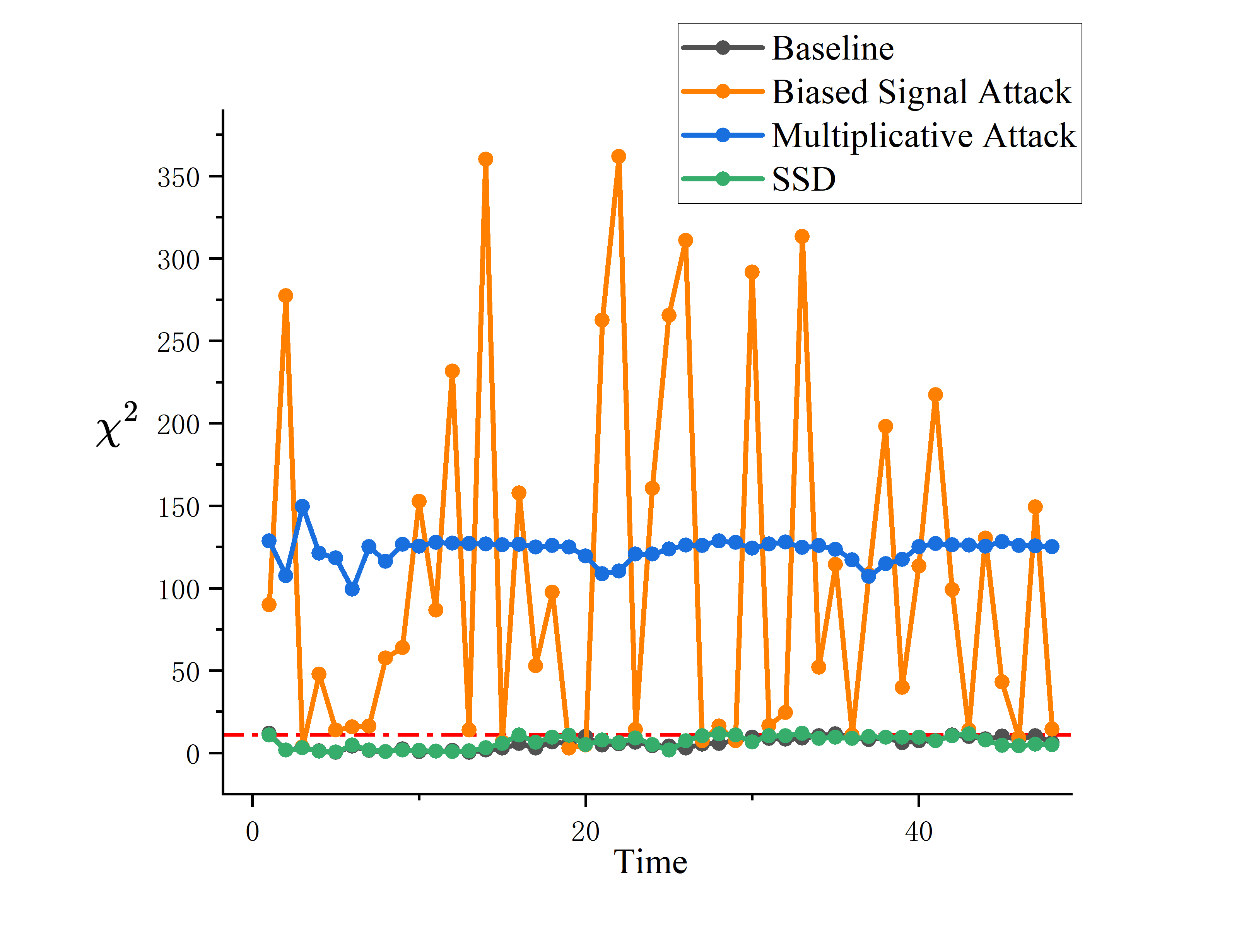}
		\caption{$\chi^2 detection$ of different attack under Trajectory \uppercase\expandafter{\romannumeral2}}
		\label{fig:ste2}
	\end{subfigure}
	\centering
	\begin{subfigure}{0.325\linewidth}
		\centering
		\includegraphics[width=0.9\linewidth]{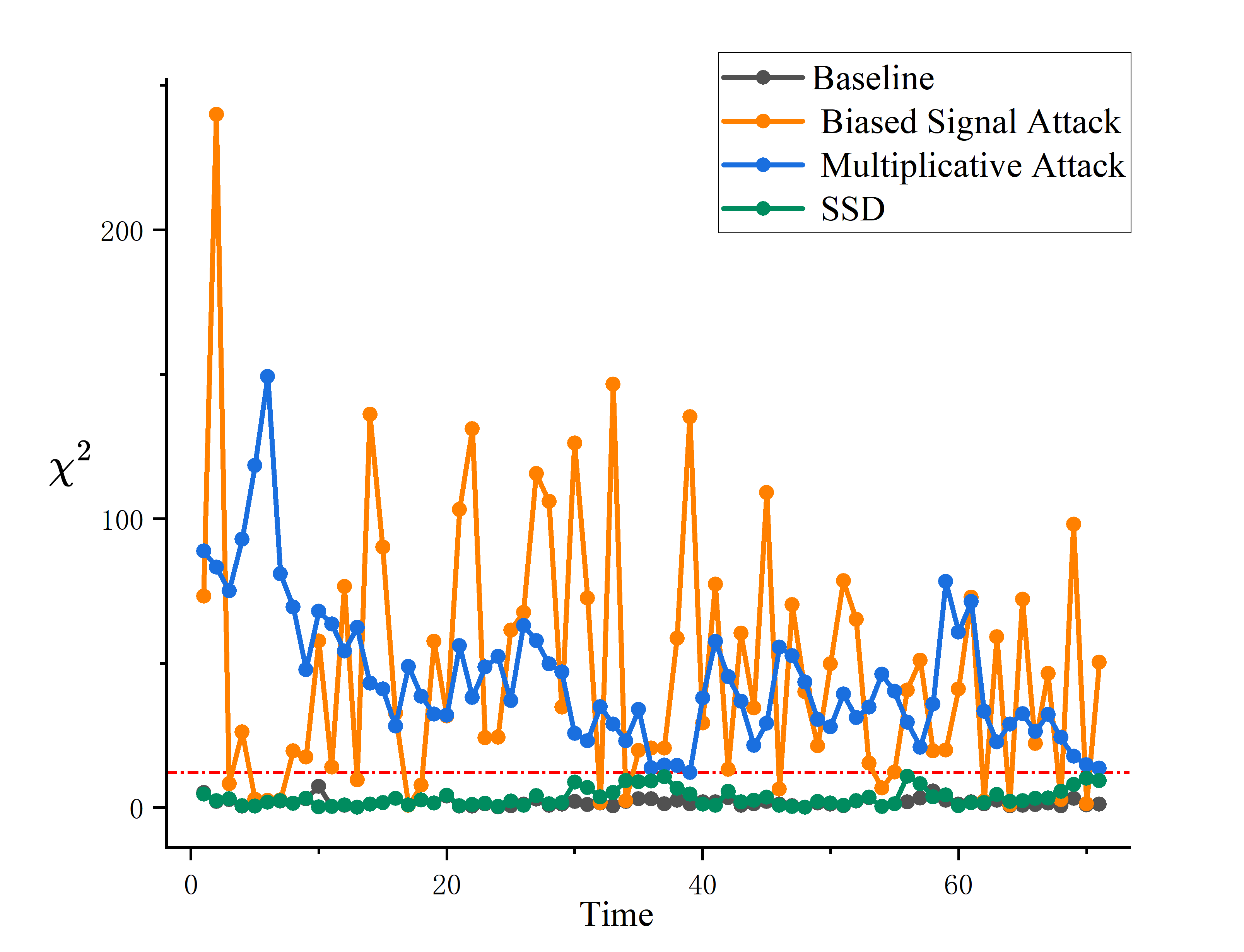}
		\caption{$\chi^2 detection$ of different attack under Trajectory \uppercase\expandafter{\romannumeral3}}
		\label{fig:ste3}
	\end{subfigure}

	\caption{Attack Stealthiness Evaluation}
	\label{fig:stealthiness}
\end{figure*}

To validate the stealthiness, we selected a chi-square detector for attack detection. For the threshold, we selected a value corresponding to a 95\% confidence level, which is 11.1. We also chose biased signal and multiplicative attacks for comparison and evaluated them across three mission trajectories. Based on prior research, we designed the following attack payloads: (1) GNSS positions are added with a uniform distribution U(0, 0.0005); (2) GNSS positions are scaled by a factor of 1.5. The experimental results in Fig.~\ref{fig:stealthiness} demonstrate that SSD successfully limits detection statistics to the threshold range and bypasses both detection methods with a carefully chosen set of attack parameters. 
Notably, compared to these works \cite{tang2023gan,fei2020learn}, we significantly reduced the loadings applied for the biased signal Attack. However, both detection methods were able to detect the attack effectively. It is clear that SSD exhibits strong stealthiness properties under chi-square detection methods and several motion states.

\begin{figure*}[htbp]
	\centering
	\begin{subfigure}{0.325\linewidth}
		\centering
		\includegraphics[width=0.9\linewidth]{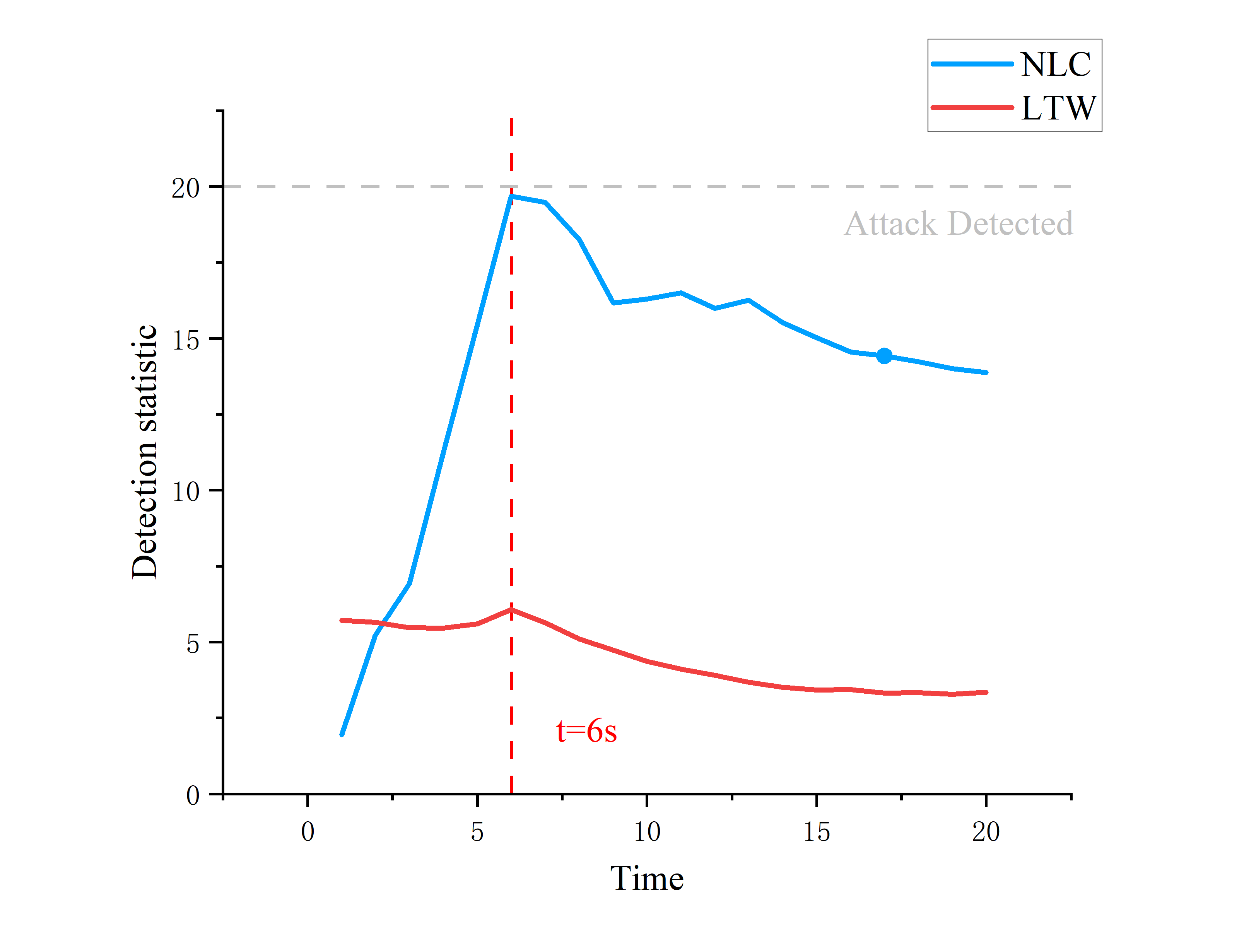}
		\caption{Detection statistic in Trajectory \uppercase\expandafter{\romannumeral1} }
		\label{fig:nlc1}
	\end{subfigure}
	\centering
	\begin{subfigure}{0.325\linewidth}
		\centering
		\includegraphics[width=0.9\linewidth]{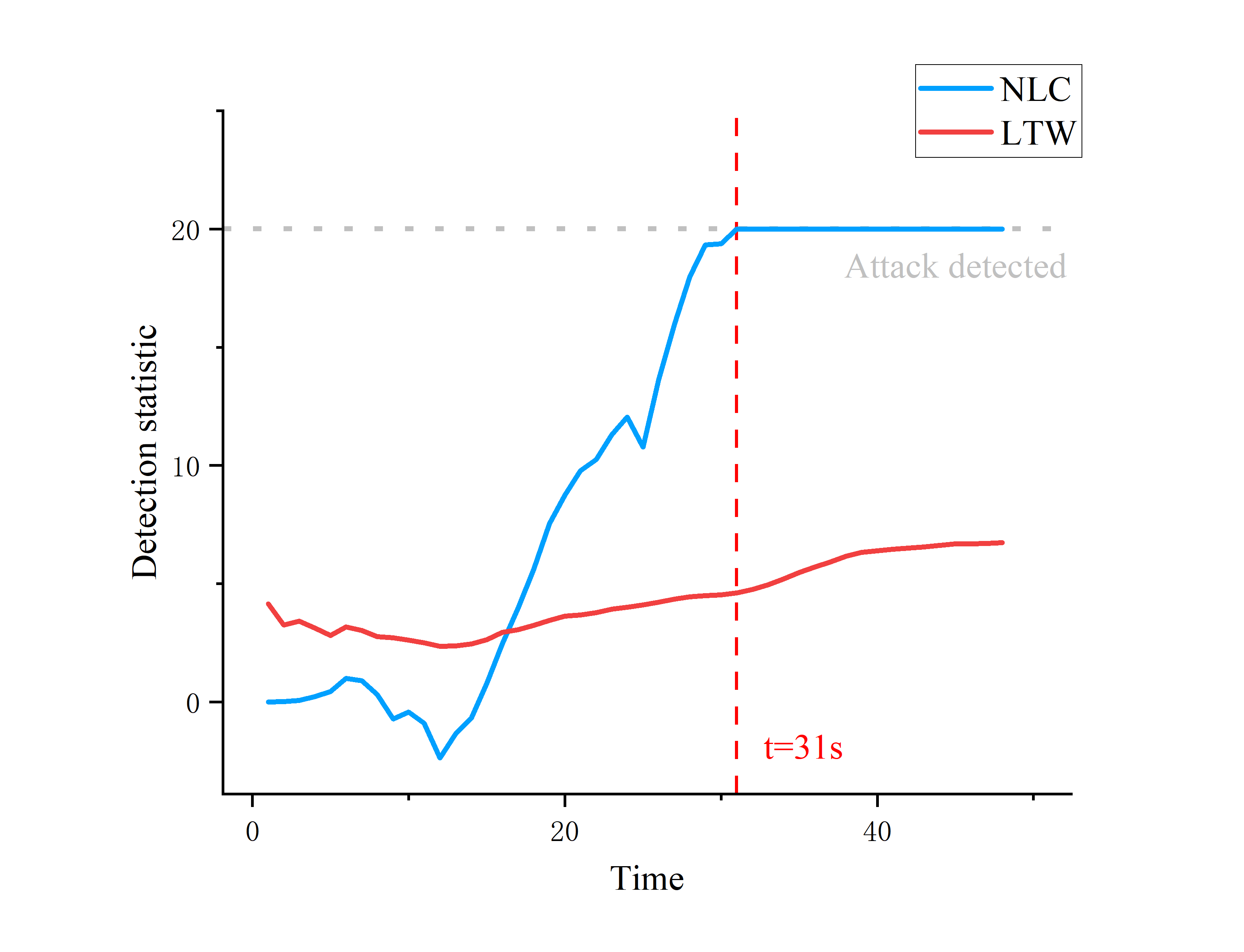}
		\caption{Detection statistic in Trajectory \uppercase\expandafter{\romannumeral2}}
		\label{fig:nlc2}
	\end{subfigure}
	\centering
	\begin{subfigure}{0.325\linewidth}
		\centering
		\includegraphics[width=0.9\linewidth]{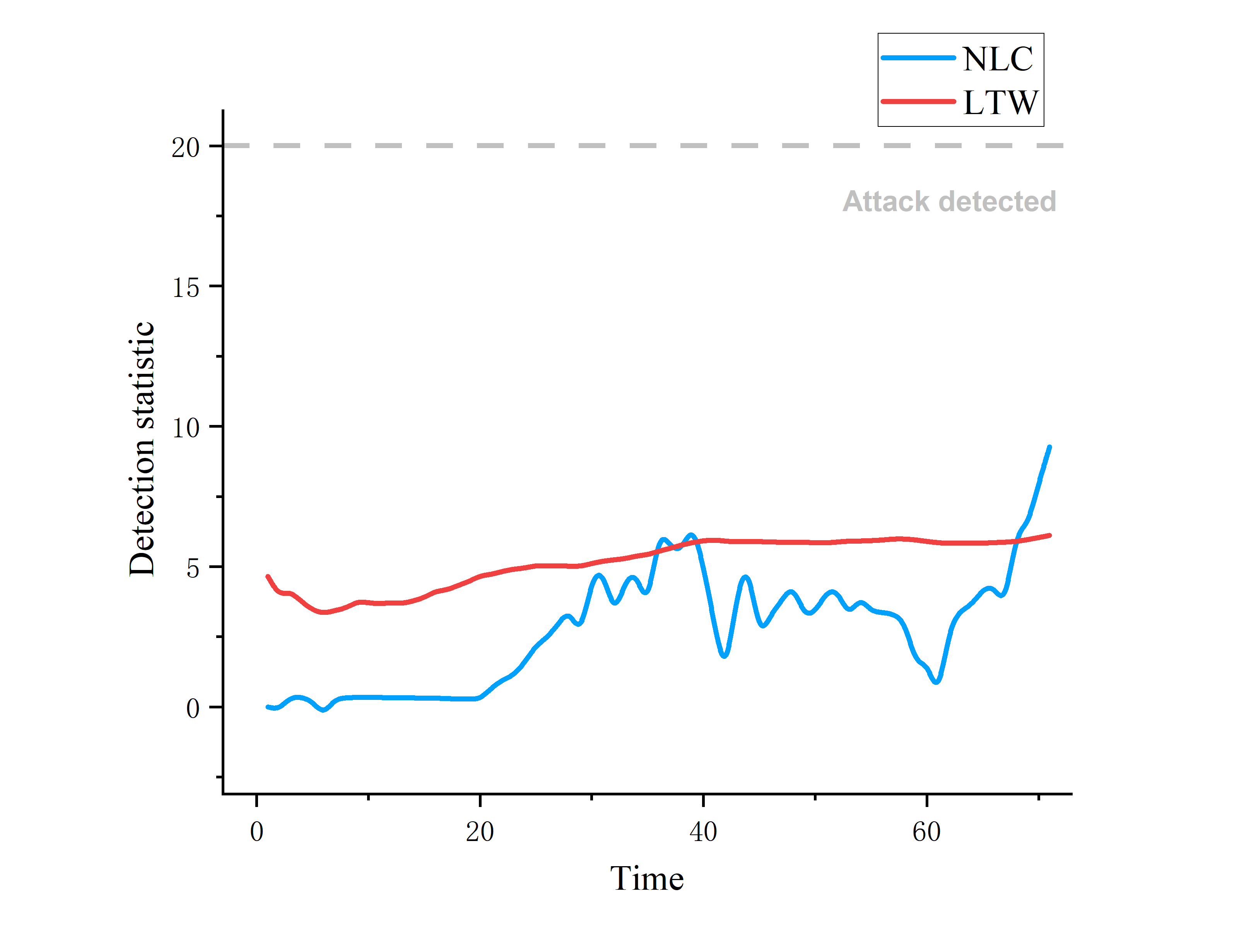}
		\caption{Detection statistic in Trajectory \uppercase\expandafter{\romannumeral3} }
		\label{fig:nlc3}
	\end{subfigure}
	\caption{Detection Statistic Comparison for
NLC and LTW under SSD.}
	\label{fig:nlc}
\end{figure*}

We also use two UAV-specific detectors, NLC~\cite{quinonez2020savior} and LTW~\cite{choi2018detecting}, as baselines for comparison. These methods are highly effective at detecting GNSS attacks, and they maintain both strong accuracy and response time, even when identifying targeted stealthy attacks. We follow the threshold in previous work~\cite{quinonez2020savior}. Fig.~\ref{fig:nlc} demonstrates that in the LTW test group, the detection statistic consistently remains below the threshold boundary, indicating the superior stealth characteristics of SSD attacks. Notably, NLC statistics demonstrate the following three characters (shown in Table~\ref{steal}): 
\begin{itemize}
    \item In Trajectory \uppercase\expandafter{\romannumeral1} (pure linear motion), the detection statistic approaches the threshold at $6s$ but never exceeds the threshold.
    
    \item In Trajectory \uppercase\expandafter{\romannumeral2} (fully nonlinear motion), the detection statistic surpasses the threshold after $31s$ cumulative duration. The detection latency is largely increased compared to the previous work (about $0.3s$)~\cite{quinonez2020savior}.

     \item In Trajectory \uppercase\expandafter{\romannumeral3} (hybrid motion mode), no significant statistical fluctuations occur during $0-20s$ linear phase, with limited oscillations (peak statistic is 9.27) emerging post nonlinear component introduction at 20s.
\end{itemize}

\begin{table}[htb]
 \centering \caption{ Detection Performance Comparison for NLC}
 \label{steal}
\begin{tabular}{cccc}
    \toprule[1.5pt]
      & \textbf{Detection Latency(s)} & \textbf{Peak Statistic} &\textbf{Success Rate} \\
    \midrule[1pt]
    
     Trajectory \uppercase\expandafter{\romannumeral1} & N/A& 19.69 & 0\%\\\hline
     Trajectory \uppercase\expandafter{\romannumeral2} & 31s & \textgreater $20$& 35.4\%\\\hline
    
    Trajectory \uppercase\expandafter{\romannumeral3}& 
    N/A& 9.27& 0\%  \\
    \bottomrule[1.5pt]
     \end{tabular}
\end{table}
These findings demonstrate significant motion-dynamic sensitivity disparities in NLC detectors. However, when attackers combine motion-state transition strategies,
positional attacks can realize residual normalization during linear phases and reduce detection sensitivity
velocity attacks to counteract statistical fluctuations from nonlinear components. These two methods make SSD show greater stealthiness in hybrid motion missions.

\subsection{Attack Effectiveness}
We start by determining the optimal combination of $Q_k$ and $R_k$ for each scenario through offline learning. With these optimal values, we proceed to validate the effectiveness of the SSD. For each combination of INS and trajectory, the UAV is tasked with following the designated path to complete a full mission, while the SSD is deployed to attack the flight. Table \ref{ae} presents the changes in each metric before and after the attack. On average, ADE/FDE is increased by 425\%/591\%. The lateral (N)/longitude (E) deviation
 reaches 3.54/3.46 meters. We will analyze the factor based on the experiment on three scenarios.

\begin{table*}[htb]
 \centering \caption{Attack Effectiveness}
 \label{ae}
\begin{tabular}{c|c|c|c|c|c}
    \toprule[1.5pt]
    \textbf{\multirow{2}{*}{Model}} & \textbf{\multirow{2}{*}{Scenario}}& \textbf{\multirow{2}{*}{Duration}} & \textbf{ADE} & \textbf{FDE} & \textbf{APDE}\\
    \cline{4-6}
    
    &&&\textbf{Normal/Attack (meters)}&\textbf{Normal/Attack (meters)}&\textbf{Normal/Attack (meters)} \\
        \midrule[1pt]
   	ECL EKF2 &Trajectory \uppercase\expandafter{\romannumeral1} & \multirow{2}{*}{20s}& (0.95)/(5.68)  & (1.80)/(5.89) & (0.89)/(5.66) 
    \\\cline{1-2} \cline{4-6}
    
    CD-EKF  &Trajectory \uppercase\expandafter{\romannumeral1} & & (1.62)/(4.64)  & (1.49)/(7.54)  &  (1.81)/(4.92)  \\\hline

     ECL EKF2  & Trajectory \uppercase\expandafter{\romannumeral2}& \multirow{2}{*}{48s}& (1.31)/(4.09)  & (1.78)/(6.15) &  (1.48)/(4.11) \\
     \cline{1-2} \cline{4-6}
      CD-EKF & Trajectory \uppercase\expandafter{\romannumeral2} & & (1.39)/(3.97)   &(1.99)/(6.47)   & (2.35)/(5.61)   \\
     \hline
      ECL EKF2 & Trajectory \uppercase\expandafter{\romannumeral3}& \multirow{2}{*}{71s} & (1.46)/(5.38)  & (0.95)/(10.47)  & (1.86)/(6.49)   \\
     \cline{1-2} \cline{4-6}
      CD-EKF & Trajectory \uppercase\expandafter{\romannumeral3}& &(1.86)/(6.35)  & (1.34)/(7.90)  & (1.66)/(6.73)  \\ 
     \bottomrule[1.5pt]
     \end{tabular}
\end{table*}

\begin{figure*}[htbp]
	\centering
	\begin{subfigure}{0.325\linewidth}
		\centering
		\includegraphics[width=0.9\linewidth]{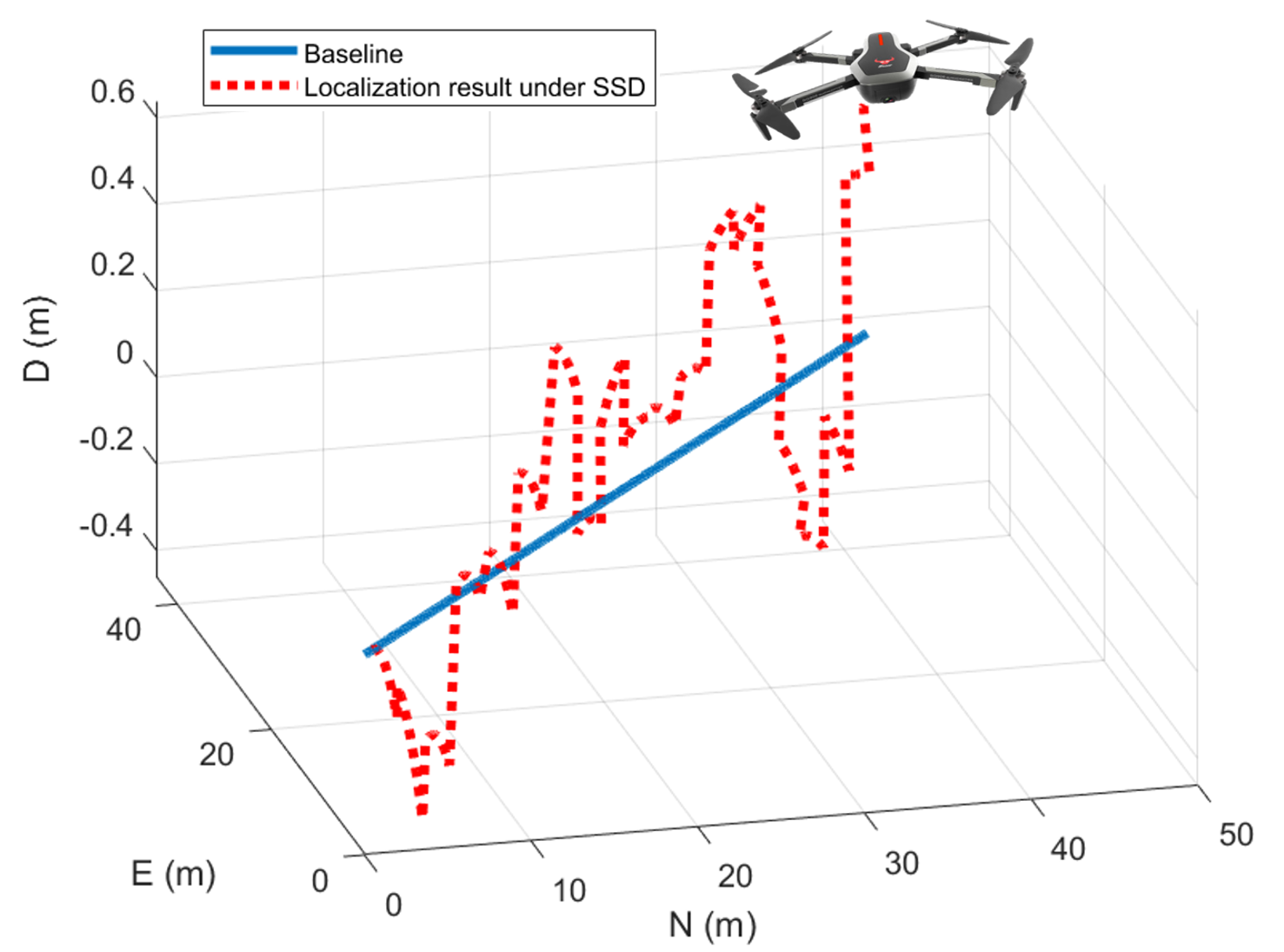}
		\caption{Attack visualization in Trajectory \uppercase\expandafter{\romannumeral1}}
		\label{fig:v1}
	\end{subfigure}
	\centering
	\begin{subfigure}{0.325\linewidth}
		\centering
		\includegraphics[width=0.9\linewidth]{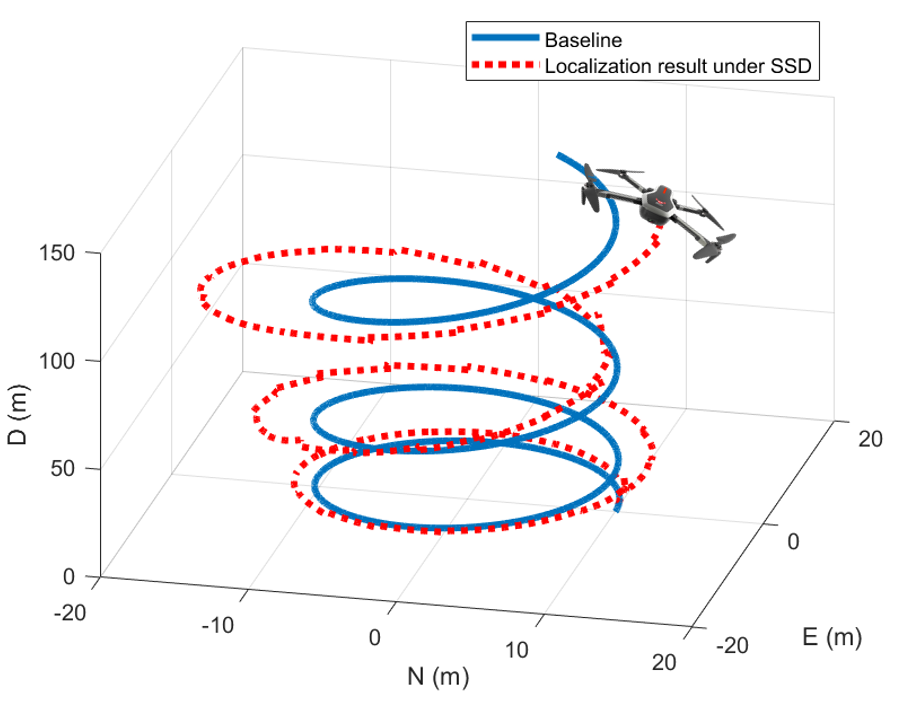}
		\caption{Attack visualization in Trajectory \uppercase\expandafter{\romannumeral2}}
		\label{fig:v2}
	\end{subfigure}
	\centering
	\begin{subfigure}{0.325\linewidth}
		\centering
		\includegraphics[width=0.9\linewidth]{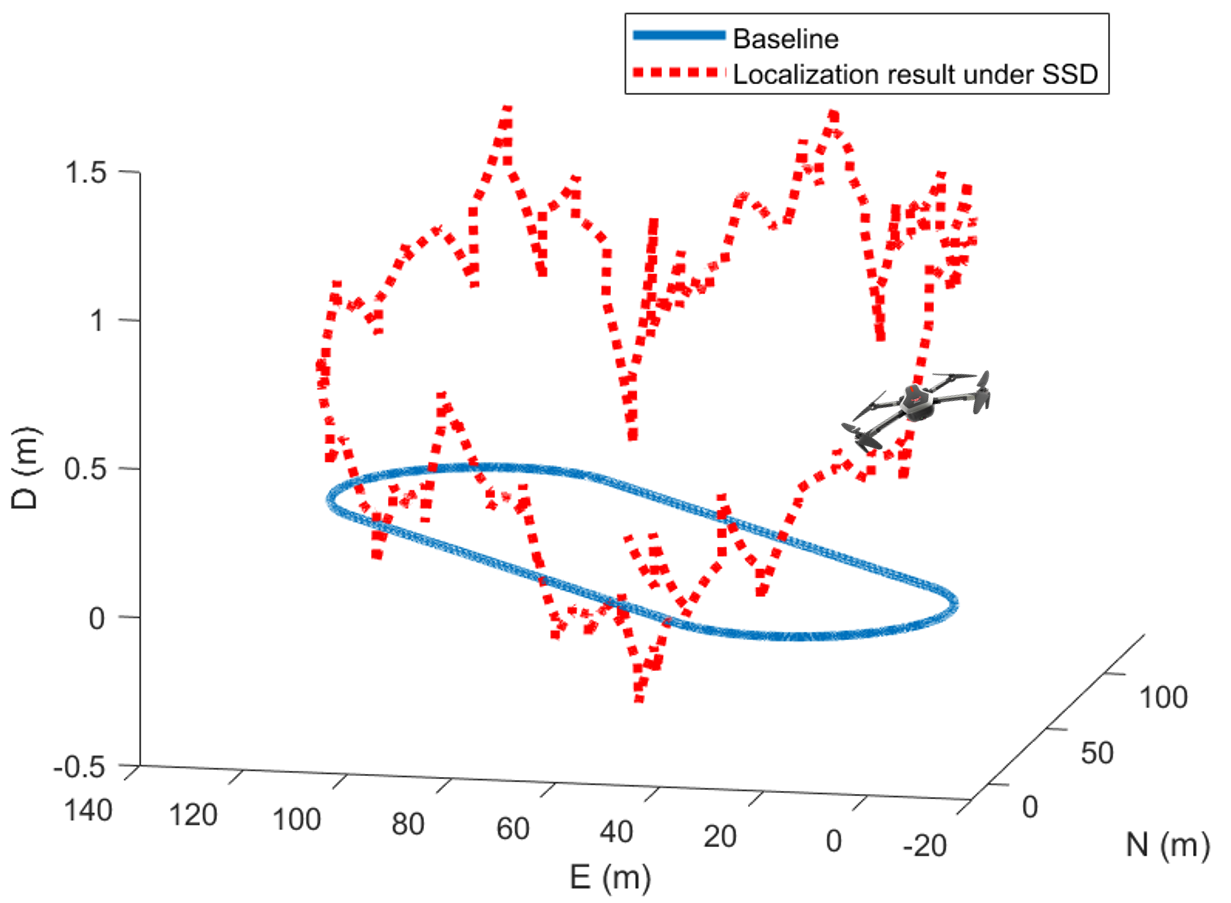}
		\caption{Attack visualization in Trajectory \uppercase\expandafter{\romannumeral3}}
		\label{fig:v3}
	\end{subfigure}
	\caption{Attack Visualization  }
	\label{fig:visualization}
\end{figure*}

\subsubsection{Different Scenarios}
In terms of scenarios, the SSD shows a greater increase in positioning error in the purely linear motion state (Fig.~\ref{fig:v1}) compared to the purely nonlinear state (Fig.~\ref{fig:v2}). This difference arises from the time-varying nature of the velocity vector in the nonlinear state, which leads to attenuation of the indirect positional interference caused by the velocity perturbation $G(t_i,a^d_i;\phi)$. However, experiments with Section \ref{section:stealthy} demonstrate that a velocity attack can still ensure the fulfillment of the stealthy precondition. When the mission involves mixed kinematic modes, the combination of velocity and positional attacks results in an additive interaction, as shown in Table \ref{ae}. The baseline increase is 366.44\% for ADE and 1102.11\% for FDE, confirming that SSD enhances attack effectiveness synergistically. Next, we will quantitatively analyze the impact of this state on mission completion rates.

\subsubsection{Impact on Mission}
When the UAV's trajectory deviation exceeds the tolerance range, the mission completion rate shows a clear decline. In three different scenarios, the APDE metrics increased by $635.96\%$, $277.70\%$, and $348.92\%$, respectively, indicating that SSDs significantly compromise mission reliability. Notably, in Scenario Trajectory \uppercase\expandafter{\romannumeral3}, where the mission duration is 71 seconds, the FDE and APDE deviations reached $10.47m$ and $6.49m$, respectively—the largest among the three scenarios. This demonstrates that SSD lowers the mission completion rate and impedes the UAV's ability to return accurately after completing the mission (shown in Fig.~\ref{fig:v3}). This effect highlights its kinematic chain impact since the mission-critical point is typically located at the motion state transition (e.g., hover → level flight, acceleration → deceleration). The attack-induced cumulative error in position estimation propagates into the next motion state, where it is further amplified.

\subsubsection{Attack Transferability}
Different INS architectures manage noise and system uncertainty in distinct ways, potentially impacting the effectiveness of SSD. To assess the cross-system adaptability of the SSD, we conducted validation experiments using CD-EKF and ECL EKF2. CD-EKF relies on continuous-time prediction with discrete-time updates, enhancing its adaptability to errors in nonlinear motion states and making its short-term error estimation more robust. However, SSD effectively exploits CD-EKF’s sensitivity to state uncertainty by inducing motion instability, thereby continuously disrupting navigation accuracy. Experimental results show that under the CD-EKF system, SSD achieves an average improvement of 304.48\% in ADE, 473.57\% in FDE, and 305.87\% in APDE (see Table \ref{ae}). These findings confirm SSD’s generalizability across EKF-based navigation frameworks.

\section{Discussion}
This study reveals the coupling mechanism between UAV motion dynamics and cyber attacks' effectiveness through systematic empirical analysis. Experimental data indicate that changes in motion state can significantly enhance the success rate of attacks. While the SSD approach demonstrates clear advantages, its engineering implementation faces two major challenges: \textbf{Dependency on A Priori Knowledge.}
The effectiveness is highly contingent upon the real-time accuracy of the object detection and tracking system. However, the existing YOLOv5 architecture, for instance, exhibits exponential decay in the Intersection over Union (IoU) metric over time in dynamic target tracking scenarios. This results in a tracking failure probability exceeding 73\% after 60 seconds of continuous locking. \textbf{Energy-Concealment Trade-off Paradox.}
While the sustained attack mode can maintain a stealthiness threshold, it leads to a non-linear increase in energy consumption on the attacking end. This escalation doesn't align with the requirements in real-world mission scenarios.

In addition, a key assumption of the SSD is that the $Q_k$ and $R_k$ are preset offline. This is a common practice in current engineering applications. From a theoretical standpoint, adaptive $Q_k$ and $R_k$ values can modify the system's sensitivity to attacks, potentially enabling the mitigation of such attacks. This insight provides a constructive direction for defending SSD. We observed that increasing $Q_k$ and decreasing $R_k$ could reduce the system's sensitivity to attacks. However, this adjustment comes at the cost of a decrease in localization accuracy. While it is possible to improve positioning accuracy by increasing $R_k$ and decreasing $Q_k$, this also makes the system more vulnerable to attacks. Consequently, a dynamic strategy for adjusting $Q_k$ and $R_k$ is crucial. This can be achieved through optimization methods or reinforcement learning, which can fine-tune these parameters in real time, balancing between accuracy and security.

\section{Conclusion}
In this paper, we investigate cybersecurity threats of UAV route planning under different motion states. We assess the effectiveness of GNSS attacks under various motion states through theoretical and experimental analyses. Our findings reveal that INS is more vulnerable during maneuvering than in linear flight. Based on this insight, we introduce SSD, a novel state-based stealthy backdoor attack, which strategically combines GNSS velocity and position attacks to exploit this vulnerability. We conducted extensive experiments, and the results show that SSD demonstrates superior effectiveness and stealthiness compared with previous methods. We hope that this work will inspire INS designers and developers to prioritize code security and implement robust dynamic defenses.


%



\section*{Acknowledgment}
This work is supported by the National Natural Science Foundation of China ( No.62233014, No.62103330), and the Innovation Foundation for Doctor Dissertation of Northwestern Polytechnical University (CX2023023).

\bibliographystyle{unsrt}
\bibliography{ref1} 
\end{document}